%% file: paper.tex
\documentclass[10pt,letterpaper,journal,cspaper,compsoc]{IEEEtran}
\usepackage{amsmath}
\usepackage{amsfonts}
\usepackage{amssymb}
\usepackage{times}

\usepackage{algorithmic}
\usepackage[ruled,vlined]{algorithm2e}
\renewcommand{\CommentSty}[1]{\textnormal{#1}}
\DontPrintSemicolon
\SetKwComment{tcp}{$\triangleright$ }{}
\SetVlineSkip{0cm}

\usepackage{graphicx}
\usepackage{wrapfig}
\usepackage{hyperref}
\usepackage{breakurl}
\usepackage{url}
\makeatletter \def\url@leostyle{\@ifundefined{selectfont}{\def\UrlFont{\sf}}{\def\UrlFont{\scriptsize\ttfamily}}} \makeatother
\renewcommand{\algorithmcfname}{ALGORITHM}%
\SetAlFnt{\small}
\SetAlCapFnt{\small}
\SetAlCapNameFnt{\small}
\SetAlCapHSkip{0pt}
\IncMargin{-\parindent}

\usepackage{colortbl}
\usepackage{multicol}

\usepackage[center,font=small]{caption}
\usepackage{subcaption}
\newcommand{\subparagraph}{}
\usepackage[compact]{titlesec}

\newcommand{\comments}[1]{}

\usepackage{array}
\usepackage{amsfonts}
\usepackage{amssymb}
\usepackage[inline]{enumitem}

\begin{document}

  \title{Topic-Based Influence Computation in Social Networks under Resource Constraints}

  \author{
    Kaan~Bing\"{o}l,
    Bahaeddin~Eravc{\i},
    \c{C}a\u{g}r{\i}~\"{O}zgen\c{c}~Etemo\u{g}lu,
    Hakan Ferhatosmano\u{g}lu,
    Bu\u{g}ra~Gedik%
    \IEEEcompsocitemizethanks
    {
      \IEEEcompsocthanksitem K. Bingol, B. Eravc{\i}, H. Ferhatosmano\u{g}lu, and
        B. Gedik are with the Department of Computer Engineering, Bilkent University, 
        Bilkent, Ankara, Turkey. Contact e-mail: kbingol@icloud.com.
      \IEEEcompsocthanksitem \c{C}. \"{O}. Etemo\u{g}lu is with T\"urk Telekom, Istanbul, Turkey.
    }%
    \thanks{}
  }

  \markboth{IEEE TRANSACTIONS ON SERVICES COMPUTING,~VOL.~PP, NO.~99, DOI 10.1109/TSC.2016.2619688}%
  {Bing\"{o}l \MakeLowercase{\textit{et al.}}: Topic-Based Influence Computation in Social Networks under Resource Constraints}

  \IEEEcompsoctitleabstractindextext{
  \input{abstract}
  }

  \maketitle
  
  \input{introduction}
  \input{resourceproblem}
  \input{systemarchitecture}
  \input{dynamicdatafetching}
  \input{experiments}
  \input{futuredirections}
  \input{related}
  \input{conclusion}
  \input{acknowledgments}

  \bibliographystyle{IEEEtran} 
  \bibliography{paper}

  \input{bio}

\end{document}

%% file: abstract.tex
\begin{abstract}
As social networks are constantly changing and evolving, methods to analyze
dynamic social networks are becoming more important in understanding social
trends. However, due to the restrictions  imposed by the social network
service providers, the resources available to fetch the entire contents of a
social network are typically very limited. As a result, analysis of dynamic
social network data requires maintaining an approximate copy of the social
network for each time period, locally. In this paper, we study the problem of
dynamic network and text fetching with limited probing capacities, for
identifying and maintaining influential users as the social network evolves.
We propose an algorithm to probe the relationships (required for global
influence computation) as well as posts (required for topic-based influence
computation) of a limited number of  users during each probing period, based
on the influence trends and activities of the users. We infer the current
network based on the newly probed user data and the last known version of the
network maintained locally. Additionally, we propose to use link prediction
methods to further increase the accuracy of our network inference. We employ
PageRank as the metric for influence computation. We illustrate how the
proposed solution maintains accurate PageRank scores for computing global
influence, and topic-sensitive weighted PageRank scores for topic-based
influence. The latter relies on a topic-based network constructed via weights
determined by semantic analysis of posts and their sharing statistics. We
evaluate the effectiveness of our algorithms by comparing them with the true
influence scores of the full and up-to-date version of the network, using data
from the micro-blogging service Twitter. Results show that our techniques
significantly outperform baseline methods ($80\%$ higher accuracy for network
fetching and $77\%$ for text fetching) and are superior to state-of-the-art
techniques from the literature ($21\%$ higher accuracy).
\end{abstract}

%% file: introduction.tex
\section{Introduction}\label{sec:introduction}

\noindent Analysis of social networks have attracted significant research
attention in recent years due to the popularity of online social networks
among users and the vast amount of social network data publicly available for
analysis. Applications of social network analyses are abound, such as
influential user detection, community detection, information diffusion,
network modeling, user recommendation, to name a few.

Influential user detection is a key social analysis used for opinion mining,
targeted advertising, churn prediction, and word-of-mouth marketing. Social
networks are dynamic and constantly evolving via user interactions.
Accordingly, the influence of users within the network are also dynamic.
Beyond the current influence of users, tracking the influence trends provides
greater insights for deeper analysis. By combining the patterns of the past
with the current information, comprehensive analysis on customers, marketing
plans, and business models can be performed more accurately. For example,
forecasting future user influences can be used to detect `rising stars', who
can be employed in upcoming on-line advertisement campaigns.

In this paper, we address the problem of identifying and tracking influential
users in dynamic social networks under real-world data acquisition resource
limits. The current approaches for influence analysis mostly assume that the
graph structure is static, or even when it is dynamic, the data is completely
known and stored in a local database. However, in many cases, analysts are
third-party clients and do not own the data. They cannot keep the data
completely fresh as changes happen, since it is typically gathered from a
service provider with limitations on resources or even on the amount of data
provided. Third-party data acquisition tools access the data via rate-limited
APIs, which constraint the fetching capacity of clients. These externally
enforced limits prevent the collection of entire up-to-date data within a
predetermined period. To this end, we present an effective solution to
rate-limited fetching of evolving network relations and user posts. Our system
maintains a local, partially fresh copy of the data and calculates influence
scores based on inferred network and text data. The proposed solution probes
limited number of active users whose influence scores are changing
significantly within the network. By combining previous and the newly probed
network data, we are able to calculate the current user influences 
accurately. The local
network copy is maintained while consuming resources within allowed limits,
and at the same time, influence values of the users are computed as accurately
as possible.

While computing and maintaining influence scores, we consider both global and
topic-based influence. Active and influential users mostly affect the general
opinion with respect to their topics of authority. For instance, a company
marketing sports goods will be interested in locating users who have high
influence in sports, rather than the global community. While this leads us to
consider topic-based analyses in our problem setting, general influence scores
of users are still of interest as well. For instance, a politician would
prefer a broader audience and identify a list of globally influential users to
promote her cause. In our system, we utilize both global and topic-based
networks and compute global as well as topic-based influences.

To demonstrate the effectiveness of our solutions, we use
Twitter~\cite{twitter}. Twitter is a good fit for research on dynamic user
influence detection due to its large user base and highly dynamic user
activity. One can collect two-way friendship relations as well as one-way
follow, re-tweet, and favorite relations via the publicly available Twitter
APIs. These APIs have well-defined resource limits~\cite{twitterAPIlimits},
which motivates the need for our probing algorithms. We calculate
PageRank~\cite{page1999pagerank} on the Twitter network as the influence score
for the users. To generate topic-based influence scores, we adapt the weighted
PageRank~\cite{xing2004weighted}, and adjust the initial scores and transition
probabilities based on topic relevance scores of the users. The topic
relevance scores are computed based on user posts, using text mining
techniques, as well as the re-tweet and favorite counts of the tweets.

To further improve the accuracy of our network inference, we perform link
prediction using trends on user relationships. The proposed solution shows
increased accuracy on Twitter data when compared with other methods from the
literature. Estimated network structure is shown to be very close to the
actual up-to-date network, with respect to influential users.  The proposed
solutions address not only the limitations of data fetching via public APIs,
but also local processing when the resources are limited to fetch the entire
data. We summarize our major contributions as follows:

\begin{itemize}

\item We estimate global and topic-based influence of users within a dynamic
social network. For topic-based influence estimation, we construct topic-based
networks via semantic analyses of tweets and the use of re-tweet and favorite
statistics for the topic of interest.

\item We propose efficient algorithms for collecting dynamic network and text
data, under limited resource availability. We leverage both latest known user
influence values, as well as the past user influence trends in our probing
strategy. We further improve our probing techniques by applying link
prediction methods.

\item We evaluate our proposed algorithms and compare results to several
alternatives from the literature. The experimental results for 
relationship fetching used for influence estimation show
that the proposed algorithms perform $80\%$ better than the baseline methods,
and $21\%$ better than the state-of-the-art method from the literature in
terms of mean squared error. For tweet fetching methods used for topic-based
influence detection, our algorithms perform $77\%$ better than the alternative
baselines in terms of the Jaccard similarity measure. \end{itemize}

The rest of this paper is organized as follows.
Section~\ref{sec:limitationproblem} describes the resource constraint problem
for data collection. Section~\ref{sec:systemarchitecture} gives the
overall system architecture and presents influence estimation techniques.
Section~\ref{sec:dynamicdatafetching} explains algorithms and strategies
proposed for the network and text fetching problems.
Section~\ref{sec:experiments} discusses results obtained from experiments run
on real data. Section~\ref{sec:related} discusses related work.
Section~\ref{sec:conclusion} concludes the paper.

%% file: resourceproblem.tex
\section{Problem Definition}\label{sec:limitationproblem}

\noindent
Our goal is to determine top-m influential users in the network, under a
constrained probing setting. Among various methods to calculate a user's
influence in the network, we have chosen PageRank based methods, since PageRank
is well understood and used
widely in the literature for various network structures. While computing
influence, PageRank naturally considers the number of followers a user has,
but more importantly it takes into account the topological place of the user
within the network. Therefore, we assume that a user's influence in the
network corresponds to its PageRank score. As a result, the top-m influential
user determination problem turns into identifying the top-m users with the
highest PageRank scores. One can also utilize other approaches that can outperform 
PageRank for estimating social influence within our framework. These approaches 
need to produce a single score that will be calculated periodically for every user.

PageRank score calculation requires having access to all the relationships
present between the users of the network. This means that we need to have the
complete network data to compute exact PageRank scores. Moreover, if the
network is dynamic, the calculation needs up-to-date network data for each
time step in order to perform accurate influence analysis.

Our system continuously collects social network data (relations, tweets,
re-tweets, etc.) via the publicly available Twitter API. Twitter enforces
certain limitations on data acquisition using the Twitter APIs. There are
different limitations for different types of data acquisition requests:

\begin{itemize}

\item \emph{Relations\footnote{For the relations, Twitter
  provides two different APIs: one for fetching the user IDs for every user
  following a specified user, and another for fetching the user IDs for every
  user a specified user is following. Our system utilizes both APIs, however
  for brevity of the rate limit calculations details are omitted.}}: 15 calls
  per 15 minutes, where each call is for retrieving a user's relations.
  Moreover, if the user has more than $5K$ followers, we need an extra call
  for each additional $5K$ followers. This   means that we can update
  relations with a maximum rate of $1$ user per minute ($R_{rel} = 1$
  user/min).

\item \emph{Tweets}: 180 calls per 15 minutes, where each call is for
  retrieving a user's tweets. Moreover, if the user has more than $200$
  tweets, we need an extra call for each additional $200$ tweets. This means
  that we can update tweets with a maximum rate of $12$ users per minute
  ($R_{twt} = 12$ user/min).\footnote{the best case, if all users have 
  $\leq200$ tweets on their timelines}.

\end{itemize}

Assuming that we update the network with a period of $P$ days, we need the
following condition to hold, in order to be able to capture the entire
network of relations:
\begin{equation}\label{limitationRelFormula}
  \mbox{Number of Users} \leq R_{rel} \cdot P \cdot 1440 
\end{equation}

For getting the recent tweets of the users, we need:
\begin{equation}\label{limitationTextFormula}
  \mbox{Number of Users} \leq R_{twt} \cdot P \cdot 1440  
\end{equation}

One can easily calculate that for a network as small as $250K$ users, we need
$174$ days to update the complete network in the best case\footnote{if
all users have $\leq5$K followers, requiring a single call per user.}.
This analysis shows that the rate limits hinder the timeliness of the data
collection process, which in turn affects the timeliness of the calculation
process to find and track influential users in the network. Furthermore,
Twitter is a highly dynamic network that evolves at a fast rate, which means
that refreshing the network infrequently will result in significant
degradation in the accuracy of the influence scores. Current resource limits
prohibit the system to collect the network data in a reasonable period of
time. Therefore, the evolving network's relationships and the tweet sets are
not fully observable at every analysis time step.

To overcome this limitation, we propose to determine a small subset of users
during each data collection period, whose information is to be updated. This
data collection process, which does not violate the rate limits of the API, is
sufficient to maintain an approximate network with a reasonable data
collection period, while at the same time providing good accuracy for the
estimated influence scores.

We apply the concept of \emph{probing} for efficient fetching of the dynamic
network and the user tweets. We denote a network at time $t$ as
$G_t=\{V_t,E_t\}$, where $V_t$ is the set of users and $E_t\subset V_t\times
V_t$ is the set of edges representing the follower relationship within the
network. In other words, $(u,v)\in E_t$ means that the user $u\in V_t$ is
following the user $v\in V_t$. Our model uses an evolving set of networks in
time, represented as $\{G_t\mid 0\leq t\leq T\}$. However, we assume that we
have fully\footnote{The initial probing of the network can be accelerated via
the use of multiple cooperating fetchers. However, this is clearly not a
sustainable and feasible approach for continued probing of the network, as it
requires large number of accounts, which are subject to bot detection and
suspension.} observed the network only at time $t=0$. $G_t$ where $t>0$, can
only be observed partially by probing. At each time period, we use an
algorithm to determine a subset of $k$ users and probe them via API calls. We
then update the existing local network with the new information obtained from
the probed users. In effect, we maintain a partially observed network
$G^{'}_t$, which can potentially differ from the actual network $G_t$.
Larger $k$ values bring the partial network $G^{'}_t$
closer to the actual network $G_t$. However, using large $k$ values is not
feasible due to rate limits outlined earlier. Our probing strategy should
select a relatively small number of users to probe, so that the data
collection process can be completed within the period $P$ (as determined by
Eq.~\ref{limitationRelFormula}). Furthermore, these probed users should bring
the most value in terms of performing accurate influence detection.

\textit{Dynamic Network Fetching Problem Definition}:
We assume that complete network information is available only at time $0$,
i.e., $G_{0}$ is known.  The problem is defined as determining a subset of
users of size $k$ at time $t$ (where $t\geq 1$), denoted by $U_t^N\subset V_t$ s.t. $|U_t^N|=k$, by
analyzing the local graph $G'_{t-1}$. The system will retrieve the partial graph related with $U_t^N$, which is denoted as $G_t^{p}(U_t^N)=(V_t^{p},E_t^{p}) \textnormal{ where } V_t^{p} = U_t^N $, and update the relationships
of the users included in this subset to construct the local network at time $t$, that is $G'_t$. We define the additions and deletions to the network as $\Sigma(U_t^N) = G_{t-1}' \setminus G_t^{p}(U_t^N)$ and $\Delta(U_t^N) = G_t^{p}(U_t^N)  \setminus G_{t-1}'$, respectively. Using these definitions we can find the network at time $t$, as $G_t' = G_{t-1}' \cup \Sigma(U_t^N) \setminus \Delta(U_t^N)$.

We aim to choose $U_t^N$ such that the influence scores of the estimated network $G'_t$ will be as close as possible to the true scores of the real network $G_t$. We summarize the problem as follows:
\begin{eqnarray*}
argmin_{U_t^N}(Influence(G'_t)-Influence(G_t)) \\
\textnormal{where }  G_t' = G_{t-1}' \cup \Sigma(U_t^N) \setminus \Delta(U_t^N)
\end{eqnarray*}

The final objective is to estimate the PageRank scores $PR'_v(t),
\forall v \in G_t$ as accurately as possible, using partial knowledge about $G_{t-1}$, that is $G'_{t-1}$, since we have used Pagerank as the indication of influence in this study.
\par


\textit{Dynamic Tweet Fetching Problem Definition}:
Given the tweets $T_{0}$ of all users in the network at time $0$,
the problem is defined as determining a subset of
users of size $k$ at time $t$ (where $t\geq 1$), denoted by $U_t^T\subset V_t$ s.t. $|U_t^T|=k$, by
analyzing the tweet set $T'_{t-1}$ and the local graph $G'_{t-1}$. The system will retrieve the partial tweet set for $U_t^T$, which is denoted as $T_t^{p}(U_t^T)=(V_t^{p},E_t^{p}) \textnormal{ where } V_t^{p} = U_t^T $, and update the tweet sets
of the users included in this subset  to construct the tweet set at time $t$, that is $T'_t$.

In this paper, we mainly focused on effective ways of
handling edge additions and removals. However, node changes are also
dynamically happening in the social network. The system handles node changes
by periodically renewing the seed list\footnote{this period is a configuration
that can be adjusted by a system administrator.}. For brevity and in order to
focus on the more prominent issue of edge additions and removals, seed list
updates are not performed as part of our experiments.

%% file: systemarchitecture.tex
\begin{figure}[h!t]
\centering
\includegraphics[width=\linewidth, keepaspectratio]{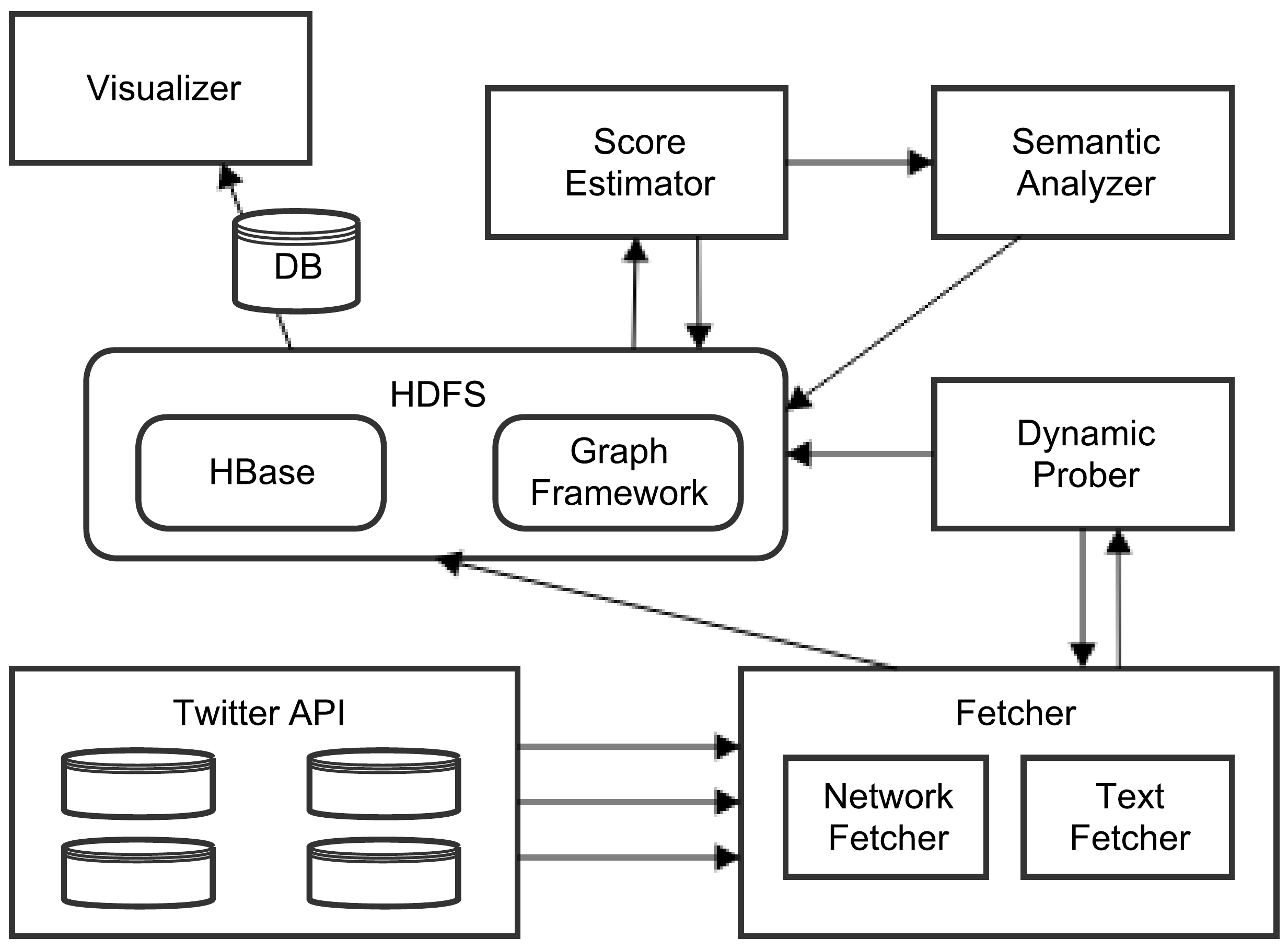}
\caption{Overall system architecture.}\label{fig:block-diagram}
\end{figure}
\section{Overall System Architecture}\label{sec:systemarchitecture}

\noindent In this section we briefly describe our system architecture, which
depicted in Figure~\ref{fig:block-diagram}.

\subsection{Social Network Data Collection}

\noindent We use the Twitter network and tweets to analyze user influence. A
Twitter network is a directed, unweighted graph where the nodes represent
users and the edges denote follower relationships in Twitter.
When a user $u$ follows another user $v$, $u$ can see what $v$ is posting, and
thus $v$ is considered to have an influence on $u$. Moreover, the user $u$ also would
have an effect on $v$'s influence, since the number of  people $v$ reaches
would potentially increase. This interaction has an effect on both users'
influence scores. In order to construct our network, we first determine a
small set of users called the \textit{core seeds}. For illustration, we
started with some popular Turkish Twitter accounts including newspapers, TV
channels, politicians, sport teams, and celebrities. Second, we collect one-
hop relations of the core seeds and add the unique users to a set called the
\textit{main seeds}. We iterate once more to collect one-hop relations of the
main seeds with a filter to avoid unrelated and inactive users. This filter
has three conditions: \begin{enumerate*}[label=\itshape\alph*\upshape)] \item
a user must have at least five followers, \item a user must have at least one
tweet within the last three months, and \item the tweet language of a user
must be Turkish. \end{enumerate*} As a result of this process, we have
determined our \textit{seed users} set, which includes approximately $2.8$
million unique users. In the final step of the data collection phase, we
acquire the relations of the seed users to determine $G_0$, that is the social
network graph at time $0$. Furthermore, we collect tweets of the seed users in
order to construct $T_0$, that the tweet set at time $0$.

We implemented the proposed methods using a distributed system with HBase and
HDFS serving as the database and file system backends. The system consists of
six main parts:
\begin{enumerate*}[label=\itshape\alph*\upshape)] 
\item local copy of the social network data on HDFS, 
\item data fetcher, 
\item dynamic prober, 
\item score estimator,
\item semantic analyzer, and 
\item visualizer.
\end{enumerate*} 
Data fetcher component, as the name implies, fetches the data (network
relations and tweets) via rate-limited Twitter APIs, periodically. Dynamic
prober makes a dynamic probing analysis, decides which users are going to be
fetched and notifies data fetcher to bring the information, accordingly. Score
estimator calculates users' influence and the related parameters of the
proposed algorithms, which are essential parts of the probing method. Semantic
analyzer performs keyword extraction and calculates the related parameters for
constructing topic-based networks. Finally, visualizer provides a graphical
user interface for result analysis.

\subsection{Score Analysis}
\noindent We calculate influence scores of users based on their 
relationships and the overall impact of their tweets in the network. We
analyze topic activities of the users from their tweets and determine
topic-based user influence scores. Overall, we are using two types of scores,
namely \textit{global influence} and
\textit{topic-based influence}, which can be interpreted together for a more
detailed analyses.
\medskip
\\
\noindent\textbf{Global Influence Score}. This score is a measure of the 
user's overall influence within the network. For this purpose we use the
PageRank ($PR$) algorithm. PageRank value $PR_v(t)$ at time $t$ for a user $v
\in G_t$ directly corresponds to the global influence score of it and will be
used interchangeably throughout the paper.

\begin{figure}[!t]
\centering
\includegraphics[width=0.75\linewidth]{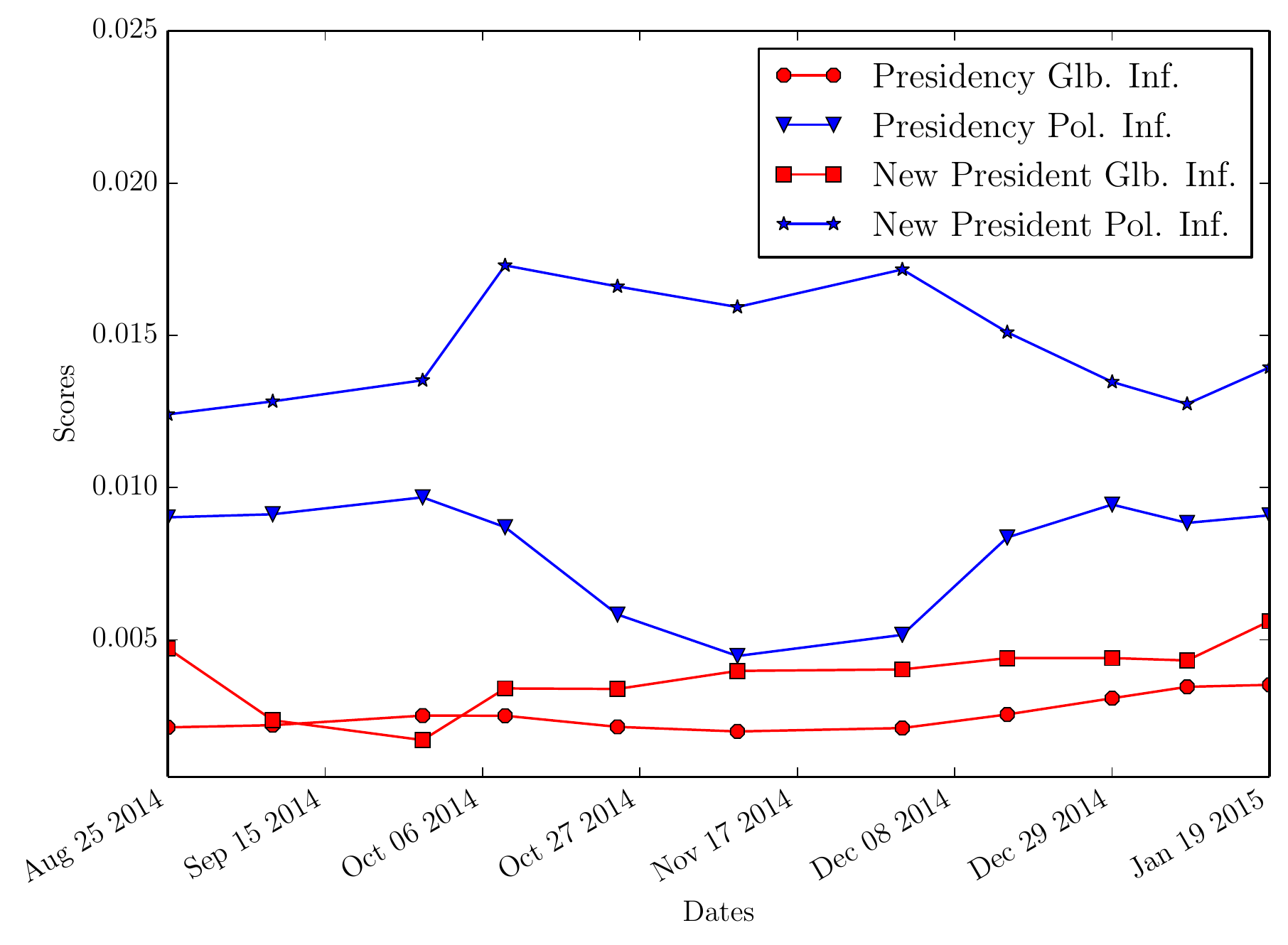}
\caption{Past global and topic-based (politics) influence scores of the 
presidency of the Republic of Turkey and the newly elected president}\label{fig:pastChanges}
\end{figure}

Figure~\ref{fig:pastChanges} illustrates the evolving nature
of the influence score by showing the global and topic-based influence scores
(calculated on true snapshots) history of users, which are selected by our
algorithm as one of the most important users that should be probed. These are
the official accounts of the presidency of the Republic of Turkey and the
newly elected president. Besides their high impact, we observe that their
influence also varies significantly over time, which further justifies the
need to probe these accounts frequently. A reason of the variation in
influence score is that the time period shown in the figure matches with the
elections for the Presidency (10 August 2014). After becoming the new
president, the president account's global influence has further increased.
During this period, it is always selected as a top user to be probed by our
proposed approach. This is intuitive, as it is a popular account with changing
influence scores over time. We can also observe the impact of presidential
change on the presidency account. During this change, its global score
slightly decreases and then starts  to increase.
\medskip \\ 
\textbf{Topic-Based Influence Score}. The system calculates topic-based
influence scores representing user activity and impact on a specific topic. We
perform semantic analysis on user tweets by taking re-tweets and favorite
counts into consideration as well. A re-tweet (RT) is a re-posting of someone
else's tweet, which helps users quickly share a tweet that they are influenced
by or like. A favorite (FAV) is another feature that represents influence
relation between users, wherein a user can mark a tweet as a favorite. These
two features help estimate the influence of an individual tweet. Since Twitter
is a micro-blogging platform, users are generally tweeting on specific topics.
While many tweets are mostly conversational and reflect self-
information~\cite{Naaman:2010,analytics2009twitter}, some are being used for
information sharing, which is important in harvesting knowledge. RTs and FAVs
are effective in separating relevant and irrelevant tweets. Accordingly, we
use them in our topic weight analysis to estimate influence of a tweet on a
specific topic.

Topic-based network construction process consists of three main phases:
\begin{enumerate*}[label=\itshape\alph*\upshape)] \item keyword extraction on
tweets, \item correlation of keywords with topic dictionaries, and \item
weight calculation.\end{enumerate*}

In the first phase, keywords are extracted from tweets by using
information retrieval techniques, including word stemming and stop word
elimination. The output from this phase is a keyword analyzed tweet corpus for
each individual user and the related histogram which captures the frequencies
of the related keywords ($K$). These corpora are further analyzed in the
second phase.

We have created a keyword dictionary ($D_j$) for each topic ($C_j$), in order
to score tweets against topics. Each dictionary contains
approximately 90 to 130 words. In order to create a dictionary for a topic, we
first compose a representative word list for the topic. We then divide these
words into groups according to context similarity and assign weights to word
groups within a scale (such as in range$[1\ldots 10]$). Context similarity can
be determined by a domain expert utilizing knowledge about the taxonomy.
Similarly, we repeat the process for all topics. As part of each dictionary,
we have assigned normalized weights to words, representing their topic
relevance. In the second phase, using the weights from the dictionaries and
the users' keyword histograms, we obtain the normalized raw topic scores of
users for each one of the topics.

In the third phase, we calculate a value called the RT-FAV total for each
user, which is the summation of the number of re-tweets and favorites received
by a user's tweets. We then multiply the normalized raw topic score by the
RT-FAV total of the user, in order to find the number of RT-FAVs the user gets
on a topic of interest. The final normalized results are used as the in-edge
weights of the users on each topic, when forming the topic-based network.

Once the topic-based network construction is complete, we execute the weighted
PageRank~\cite{xing2004weighted} ($WPR$) algorithm which also considers the
importance  of the incoming and outgoing edges in the distribution of the rank
scores. The resulting  weighted PageRank values of users, denoted by
$WPR_v(t)$ at time $t$ for  $v \in G_t$, is assigned as their topic-based
influence scores.

Due to the nature of the PageRank algorithm, some of the globally influential
users also turn out to be highly influential for most or all of the topics.
These users have a lot of followers and they are also followed by some of the
influential accounts of the specific topics, which cause them to score high
for topic-based analysis as well. Therefore, they can get high topic-based
influence scores even if they do not actively tweet about the topic itself. To
eliminate this effect, we apply one more level of filtering to remove these
globally effective accounts from the topic-sensitive influence lists. In
particular, if the number of tweets a user posted that are related with the
topic at hand is less than a predefined percentage, e.g., $\%40$\footnote{Note
that a tweet can be related to zero or more topics.}, of the total number of
tweets posted by the user, then the user is discarded for that topic's score list. This
filtering process significantly reduces the noise level in the analysis.

As a result, for each topic, we construct a weighted network in which an edge
($(u, v)$) represents the amount of topic-specific influence a user ($v$) has
on a follower user ($u$). Thus, the results of weighted PageRank algorithm
gives us the overall topic-influence scores on the network.

Figure~\ref{fig:pastChanges} also shows the topic-based score
history of the official account of the presidency of the Republic of Turkey
and the newly elected president. We can see from the figure that the change in
the topic-based scores are more dramatic compared to the global scores. This
is intuitive, as the topic-sensitive scores are depending on users' tweets and
sharing statistics. A user might be very active on some weeks about a specific
topic such that her influence on the topic might increase dramatically.
Likewise, when she posts something important, it might achieve high sharing
rates. On the other hand, when she just posts regular tweets which are not
shared, her influence on the topic might decrease quickly.

%% file: dynamicdatafetching.tex
\section{Dynamic Data Fetching}\label{sec:dynamicdatafetching}
\noindent
In this section, we introduce our algorithms for probing dynamic social
networks. In order to efficiently determine a subset of vertices to probe, we
develop heuristics for both dynamic network fetching and dynamic tweet
fetching problems given in Section~\ref{sec:limitationproblem}.

Since we have chosen the PageRank score as the indicator of influence in a
social network, we analyze its change as the network evolves. PageRank value
of a specific vertex $v$ is given as follows:

\begin{equation}\label{PageRank}
  PR(v)=\alpha \sum_{\forall (u,v) \in E_{in}(v)}{\frac{PR(u)}{|E_{out}(u)|}} + 
  \frac{1-\alpha}{n},
\end{equation}
where $PR(v)$ denotes the PageRank value, $E_{in}(v)$ denotes the in-edge set,
and $E_{out}(v)$ denotes the out-edge set for $v$.

\begin{figure}[h!t]
  \centering
  \begin{subfigure}{.4\textwidth}
    \centering
    \includegraphics[width=.35\linewidth,keepaspectratio]{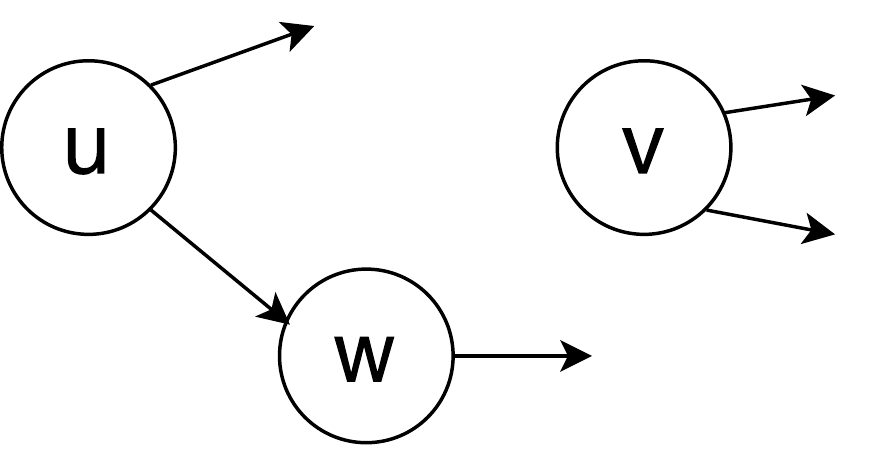}
    \caption{Previous state of the network before new edge.}
    \label{fig:SampleGraphBefore}
  \end{subfigure}
  \begin{subfigure}{.4\textwidth}
    \centering
    \includegraphics[width=.35\linewidth,keepaspectratio]{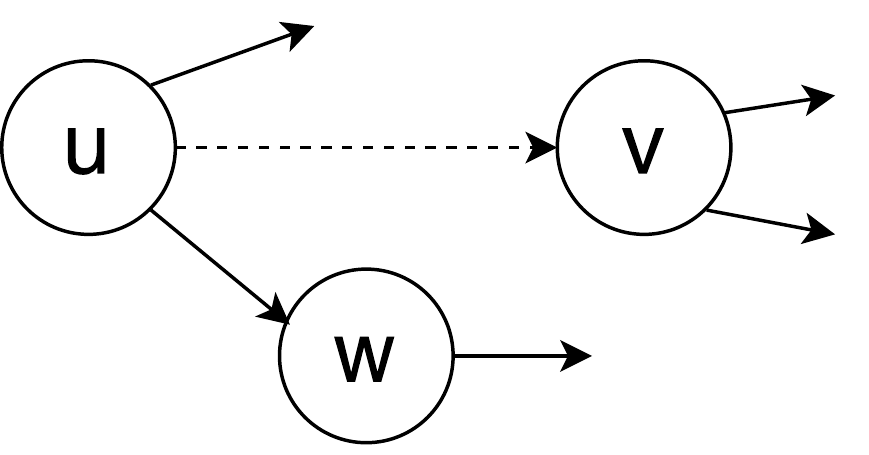}
    \caption{Current state of the network after new edge.}
    \label{fig:SampleGraphAfter}
  \end{subfigure}
  \caption{A sample network for analysis.}
  \label{SampleGraph}
\end{figure}

Figure~\ref{SampleGraph} shows an example network, which will be used to
demonstrate the effects of network changes on PageRank values.

Assume that an edge $(u,v)$ is added to the state in
Figure~\ref{fig:SampleGraphBefore} due to the evolving nature of the network.
The resulting current state is shown in Figure~\ref{fig:SampleGraphAfter}. Here,
we analyze the effect of this addition on the PageRank values of the out
neighbors of $u$. We see that the PageRank value of $v$ is as follows per
Eq.~\ref{PageRank}:

\resizebox{.98\linewidth}{!}{
\begin{minipage}{\linewidth}
\begin{align*}
  PR^{new}(v)&=\alpha \left(  \sum_{\forall (i,v) \in E_{in}(v)}{\frac{PR(i)}{|E_{out}(i)|}}+ 
  \frac{PR(u)}{|E_{out}(u)|+1}  \right) + \frac{1-\alpha}{n}  \\
  &=PR(v)+\alpha \frac{PR(u)}{|E_{out}(u)|+1}
\end{align*}
\end{minipage}}
\smallskip
\\
We can easily extend this analysis to multiple new edges since the total effect 
will be a superposition of the effect of the new individual in-edges of vertex $v$.
\resizebox{\linewidth}{!}{
\begin{minipage}{\linewidth}
\begin{align*}
PR^{new}(v)=PR(v) + \alpha \sum_{\forall (u,v)\in E_{in}^{new}(v)}{}
\frac{PR(u)}{|E_{out}(u)|+1}
\end{align*}
\end{minipage}}

PageRank values of out neighbors of $u$ other than $v$, such as $w$, are
impacted as follows:
\resizebox{\linewidth}{!}{
\begin{minipage}{\linewidth}
\begin{align*}
  PR(w)&=\alpha \left(  \sum_{\forall (i,w) \in E_{in}(w) \setminus (u,w)}
  {\frac{PR(i)}{|E_{out}(i)|}}+ \frac{PR(u)}{|E_{out}(u)|}  \right) + \frac{1-\alpha}{n}
\end{align*}
\end{minipage}}

\resizebox{.98\linewidth}{!}{
\begin{minipage}{\linewidth}
\begin{align*}
  PR^{new}(w)&= \alpha \left(  \sum_{\forall (i,w) \in E_{in}(w) \setminus (u,w)}
  {\frac{PR(i)}{|E_{out}(i)|}}+ \frac{PR(u)}{|E_{out}(u)|+1}  \right) + \frac{1-\alpha}{n}  \\
  PR^{new}(w)&= PR(w)- \alpha \frac{PR(u)}{|E_{out}(u)|.(|E_{out}(u)|+1)}
\end{align*}
\end{minipage}}
\smallskip
\\
These effects are the immediate responses on the vertices that are considered.
These residual PageRanks will ripple out to all the vertices in all the paths
from $v$ and $w$ in each iteration of the PageRank algorithm. But the effect
will decease as the residuals will be divided by the number of outgoing edges
for each vertex visited. We will analyze the effects of the first iteration of
the algorithm to simplify the problem and to get a general feel of the change
in PageRank values. Considering expected value of
$\overline{E_{out}}=E[|E_{out}(u)|]$ as the average out-degree for vertices,
the differential PageRanks are given as follows:
\begin{align}\label{PageRankDiffY}
\nabla PR(v)&= \alpha \frac{PR(u)}{\overline{E_{out}}} \\ \label{PageRankDiffZ}
\nabla PR(w)&= - \alpha \frac{PR(u)}{\overline{E_{out}}^2}
\end{align}

We can see from Eqs.~\ref{PageRankDiffY}~and~\ref{PageRankDiffZ} that we
should select the vertices, say $u$, with the following properties for
accurate $G'_t$ and $PR'_u(t)$ estimations:

\begin{itemize}
  \item vertices with high PageRank values ($PR(u)$);
  \item vertices whose PageRank values change over time;
  \item vertices with high out-degrees ($E_{out}(u)$);
  \item vertices whose out-degrees change over time.
\end{itemize}

PageRank, when computed until the values converge in steady state, considers
both incoming and outgoing edges. The parameters related to out-degree values
are intrinsically taken into account when PageRank is computed. Hence, in our
dynamic fetching approach, we focus only on PageRank values and their changes
to cover all the cases listed above.

Based on these observations, we will define a utility function that
incorporates the above findings. We will find the vertices that maximize this
utility function, which will be probed and used to estimate the influence
scores of the evolving network. We analyze two sub-problems of the general
case specific for our application: network fetching and tweet fetching. These
sub-problems and the solutions will be addressed in the subsequent sections.

\subsection{Dynamic Network Fetching using Influence Past}\label{sec:dynamicnetworkfetching}
\noindent
We aim to probe a subset, $U_t^N$, update the edges incident on vertices in
$U_t^N$ to form $G'_t$, and calculate PageRank values $PR'_v(t),\ \forall v
\in G_t$. In order to determine this subset, we use a time series of past
PageRank values for a vertex $v$, named the \emph{influence past} of $v$.
Formally, we have $IP_v= [\hdots, PR'_v(t-2), PR'_v(t-1)]$.

In our strategy for determining $U_t^N$, we consider the vertices whose PageRank
values change considerably over time. We first explored building time-series
models over sequences of scores to forecast their future values. There are
some well-known methodologies in the literature for forecasting using this
kind of time-series data, such as ARIMA models~\cite{hibon1997arma}. However, these models
typically require much longer sequences for accurate predictions. Therefore, in
order to quantify this \emph{change} for a vertex $v$, we calculate the
standard deviation of the time series $IP_{v}$, that is:
\begin{equation}\label{changeFormula}
  Change_{v}=\sigma_{IP_{v}}=\sqrt{Var(PR'_{v})}
\end{equation}

Choosing the best vertices to probe can be performed by calculating a score
that is a linear combination of the PageRank value and the change in PageRank
values, as given in Eq.~\ref{Score}. Here, $\theta$ parameter balances
the importance of the two aspects. We assume that influence past that contains
at least two data points is available for every user, in order to calculate
the score changes.%
\begin{equation}\label{Score}
  Score(v) = (1-\theta)PR'_{v}(t-1) + \theta\,Change_{v}
\end{equation}

After the selection of the users with respect to the ranking of $Score(v)$, 
we probe their current relations and form $G'_t$. 
\medskip
\\
\noindent\textbf{Round-Robin \& Change Probing.} Change Probing could cause
the system to focus on a particular portion of the network and may discard the
changes developing in other parts. This is because the probing scores of some
vertices will be stale and as a result these vertices may consistently rank
below the top-k, despite changes in their real scores. This bias could end up
accumulating errors in the influence scores of these vertices and start to
have an impact on the entire network. Therefore, we propose to use Change
Probing together with Round-Robin Probing, in which users are probed in a
random order with equal frequency. In this way, we aim to probe every vertex at
least once within a specific period $Prr$ s.t. $Prr\leq|V_t|*P/((1-\beta)*k)$.
Round-Robin Change algorithm probes some portion of the network randomly and marks 
all probed users. Thus, any probed users are not probed randomly again, until all users are probed at
least once within $P$. In this method, we control the balance between change
vs. random selection by using a parameter $\beta \in [0,1]$. In particular, we
choose $\beta*k$ users to probe with Change Probing and $(1-\beta)*k$ users
with Round-Robin Probing.
\medskip
\\
\noindent
\textbf{Network Inference.} Since we are able to fetch data only for a limited number 
of users, there is a high probability that other users in the network have
changed their connections as well. To take these possible changes into
account, we have incorporated \emph{link prediction} into our solution. Link
prediction algorithms assign a score to a potential new edge $(u,v)$ based on
the neighbors of its incident vertices, denoted as $\Gamma_{u}$ and
$\Gamma_{v}$. The basic idea behind these scores is that the two vertices $u$
and $v$ are more likely to connect via an edge if $\Gamma_{u}$ and
$\Gamma_{v}$ are similar, which is intuitive. Considering social networks, two
people are likely to be friends if they have a lot of common friends. There
are different scores used in the literature, including the common neighbors,
Jaccard's coefficient, Adamic/Adar, and Resource Allocation Index (RA). We
use RA as part of our approach, since it was found successful on a
variety of experimental studies on real-life networks~\cite{lu2011link}. One
could also adopt more advanced prediction algorithms such
as~\cite{backstrom2011supervised}, in order the increase effectiveness of this
approach.

\begin{algorithm}[t!]
  \caption{Algorithm for Dynamic Network Fetching}
  \label{Alg_Network_Fetch}
  \begin{algorithmic}
    \STATE \textbf{Input:} $G'_{t-1}$, $IP$, $PR'(t-1)$, $\theta$, $\beta \in [0,1]$, $k$, $rrRecord$
    \STATE \textbf{Output:} $G'_{t}$
    \STATE \textit{// Fetch network} 
    \FORALL{$v \in V_t$}
      \STATE $\sigma_{IP_{v}}=\sqrt{Var(IP'_{v})}$
      \STATE $Score(v)=(1-\theta)PR'_{v}(t-1) + \theta \cdot \sigma_{IP_v}$
    \ENDFOR
    \STATE $U_t^N \gets \emptyset$
    \WHILE{$|U_t^N| \leq k \cdot \beta$}
      \STATE $v \gets argmax_{v \in V_{t-1}} Score(v)$
      \STATE $U_t^N \gets U_t^N \cup \{v\}$, $V_{t-1} \gets V_{t-1} \setminus \{v\}$
    \ENDWHILE
    \WHILE{$|U_t^N| \leq k$}
      \STATE $v \gets $ randomly choose from $V_{t-1}$
      \IF{$v \notin rrRecord$}
        \STATE $U_t^N \gets U_t^N \cup \{v\}$, $V_{t-1} \gets V_{t-1} \setminus \{v\}$
        \STATE $rrRecord \gets rrRecord \cup \{v\}$
      \ENDIF
    \ENDWHILE  
    \STATE Probe $U_t^N$ for relationships, Form $G'_{t}$
    \STATE \textit{// Infer network} 
    \STATE Calculate $RA_{u,v}$, $\forall (u,v) \in \widetilde{E}=V_{t} \times V_{t}$ 
    \FOR{$E_g$ times}
      \STATE $(u,v) \gets argmax_{(u,v) \in E_t} RA_{u,v}$
      \STATE $E_{t} \gets E_{t} \cup \{(u,v)\}$
    \ENDFOR
    \STATE Output $G'_{t}$
  \end{algorithmic}
\end{algorithm}

RA is founded on the resource allocation dynamics of complex networks and gives more
weight to common neighbors that have low degree. For an edge $(u,v)$ between
any two vertices $u$ and $v$, RA is defined as follows:
\begin{equation}\label{RA_Def}
\begin{split}
  RA_{u,v}=\sum_{w \in \Gamma_{u} \bigcap \Gamma_{v}}{\frac{1}{degree(w)}},
  \\ \text{where } \Gamma_{v} \text{ is the neighbors of } v
\end{split}
\end{equation}

The $RA$ score, $RA_{u,v}$ for the edge $(u, v)$, is proportional to the
probability of an edge being formed between the vertices $u$ and $v$ in the
future. Based on this, we rank all the calculated $RA$ scores. Since the edges
in our network are not defined probabilistically and are defined
deterministically as existent or non-existent, we need to determine how many
of these scored edges should be selected. Therefore, we define a growth rate,
$E_g$, which is the average change in the number of edges ($|E|$) between
snapshots of the network after excluding the changes due to $U_t^N$. After
calculating RA scores for all possible new edges, we choose $E_g$ edges with
the highest scores. Using this method, we add new connections to the current
graph, to finally have the estimated graph $G'_{t}$. The pseudo code of the
network inference based probing algorithm we use to select $k$ vertices to
probe is given in Algorithm~\ref{Alg_Network_Fetch}.

\begin{algorithm}[t!]
  \caption{Dynamic tweet fetching via $G$-$WG$}
  \label{Alg_Network__Tweet_Fetch}
  \begin{algorithmic}
    \STATE \textbf{Input:} $T^{j'}_{t-1}$, $TIP^{j}$, $WPR^{j'}(t-1)$, $\theta$, $\beta \in [0,1]$, $k$, $rrRecord$
    \STATE \textbf{Output:} $T^{j'}_{t}$
    \FORALL{$C_j$}
      \FORALL{$v \in V^{j}_{t-1}$}
        \STATE $\sigma_{TIP_{v}}=\sqrt{Var(TIP'_{v})}$
        \STATE $Score^{j}(v)=(1-\theta)WPR^{j'}_{v}(t-1) + \theta \cdot \sigma_{TIP^{j}_v}$
      \ENDFOR
      \STATE $U^{j}_t \gets \emptyset$
      \WHILE{$|U^{j}_t| \leq k \cdot \beta$}
        \STATE $v \gets argmax_{v \in V^{j}_{t-1}} Score^{j}(v)$
        \STATE $U^{j}_t \gets U^{j}_t \cup \{v\}$, $V^{j}_{t-1} \gets V^{j}_{t-1} \setminus \{v\}$
      \ENDWHILE
      \WHILE{$|U^{j}_t| \leq k$}
        \STATE $v \gets$ randomly choose from $V^{j}_{t-1}$
        \IF{$v \notin rrRecord$}
          \STATE $U^{j}_t \gets U^{j}_t \cup \{v\}$, $V^{j}_{t-1} \gets V^{j}_{t-1} \setminus \{v\}$
          \STATE $rrRecord \gets rrRecord \cup \{v\}$
        \ENDIF
      \ENDWHILE  
      \STATE Probe $U^{j}_t$ for tweets, Form $T^{j'}_{t}$
      \STATE Output $T^{j'}_{t}$
    \ENDFOR
  \end{algorithmic}
\end{algorithm}

\subsection{Dynamic Tweet Fetching using Topic-Based Influence Past}
\label{sec:dynamictweetfetching}
\noindent
Our dynamic tweet fetching solution makes use of the weighted PageRank values
and comprises of two steps. First, we infer the evolving relationships of
the network using the methods explained earlier in the previous section. This
way we can track and estimate the changing relationships. Second, we select a
subset of users to fetch their tweet data. Specifically, we aim to probe a
subset, $U_t^T$, collect their tweets, and update the edge weights for the users
in $U_t^T$; all in order to form $WG^{j'}_{t}$ for a given topic $C_j$. We then
compute weighted PageRank values to find $WPR_{v}^{j'}(t),
\forall v \in WG_t^j$ for a given topic $C_j$. To select the subset of users
in $U_t^T$, we use a time series of the past weighted PageRank values, named the
\emph{topic-based influence past} of $v$. Formally, we have $TIP_{v}= [\hdots,
WPR^{j'}_{v}(t-2), WPR^{j'}_{v_i}(t-1)]$. This is performed independently for
all topics of interest, $\{C_j\}$ .

There are two different approaches we employ to track the topic-based influence 
scores:
\begin{itemize}
  \item Use the global network parameters for network fetching and
  the topic-sensitive network parameters for tweet fetching. This is named as the
  $G$-$WG$ method, where global $G_t$ is used for network fetching, and
  topic-sensitive $WG_t$ is used for tweet fetching.
  \item Use the topic-sensitive network parameters for both network and tweet
  fetching. This is named as the $WG$-$WG$ method.
\end{itemize}
The first approach, $G$-$WG$, is useful for cases where globally influential
users are tracked, but with minimal additional resources, topic-based
influential users are to be determined as well. This might be the only viable
option if the bandwidth is not enough for selecting and updating the vertices
separately for each topic, especially if the number of topics is high. For the
second approach, that is $WG$-$WG$, we construct separate networks $WG^j$ for
each topic and evolve them separately. We update each network at the end of a
probing period, using the new tweets fetched to track the most influential
vertices for each topic $C_j$. The high-level algorithm for the $G$-$WG$
method is given in Algorithm~\ref{Alg_Network__Tweet_Fetch}. The algorithm for
$WG$-$WG$ is very similar, and is omitted for brevity.


%% file: experiments.tex
\section{Experiments and Results}\label{sec:experiments}
\noindent
In this section, we present the experimental setup and the results of our
evaluation of the proposed algorithms. We also present experiments analyzing
the sensitivity of the parameters used.

\subsection{Data Sets}\label{sec:datasets}
\noindent
We collected data using the public Twitter API, as described in
Section~\ref{sec:systemarchitecture}. These API calls are restricted by rate
limit windows. These windows represent $15$ minute intervals and the allowed
number of calls within each window can vary with respect to the call type. Our
system makes three different calls,
\begin{enumerate*}[label=\itshape\alph*\upshape)] 
\item ``GET followers/ids'', which returns user IDs
for every user following the specified user,
\item ``GET friends/ids'', which returns user IDs
for every user the specified user is following, and
\item ``GET statuses/user\_timeline'', which returns the most recent 
Tweets posted by the specified user.
\end{enumerate*}. 
For the first two call type, we are allowed to make $15$ calls per 
window. Every call can return up to $5$K followers/friends. For the users who have more
than $5$K followers/friends, we have to make multiple calls, accordingly. For the third
type, we are allowed to make $180$ calls per window. Each call can return
$200$ tweets of the queried user. Details of the calls are also presented in
Section~\ref{sec:limitationproblem} with the accompanying analysis.

We collected the network between the end of August 2014 and the beginning
of January 2015, with a period of $15$-$20$ days. As a result, we have
obtained $11$ snapshots of the Turkish users' network with progressing
timestamps. We collected the relations of $2.8$ million users, which
amounts to a total of $310$ million edges on average. Users
are recrawled for each snapshot so that snapshots
contain exact information with respect to the network. We took the first
snapshot as the initial network to calculate the probing scores (see
Eq.~\ref{Score}) and the rest of the snapshots were used as ground truth for
the evaluation of the probing algorithms. For the topic-based influence
estimation, we also collected the tweets of our seed users in the same
period. We constructed a dataset formed of $11$ snapshots containing $5.5$
billion tweets in total. We take the first snapshot as the initial tweet set
as in the case of the relationship network analysis. From this data, we 
built up the topic weighted networks and calculated probing scores (see
Eq.~\ref{Score}), accordingly.

In our probe simulation module, we fetch the connections of the users we have
selected for probing, from the real network $G_t$ at time $t$. We then update
these connections (adding new ones and deleting old ones) on the previously
observed network $G^{'}_{t-1}$ at time $t-1$, in order to obtain the estimated
network $G^{'}_t$ at time $t$. Finally, we compare the influence estimation
results from the observed network $G^{'}_t$ with the ones from the real
network $G_t$. Same procedure is also applied for the tweet sets.

In order to include extensive number of experiments in our evaluation, we 
focused on the top $250K$ influential users and restricted the network on
which the scores are computed to the network formed by these users.

\begin{figure}[t]
\centering
\includegraphics[width=1\linewidth]{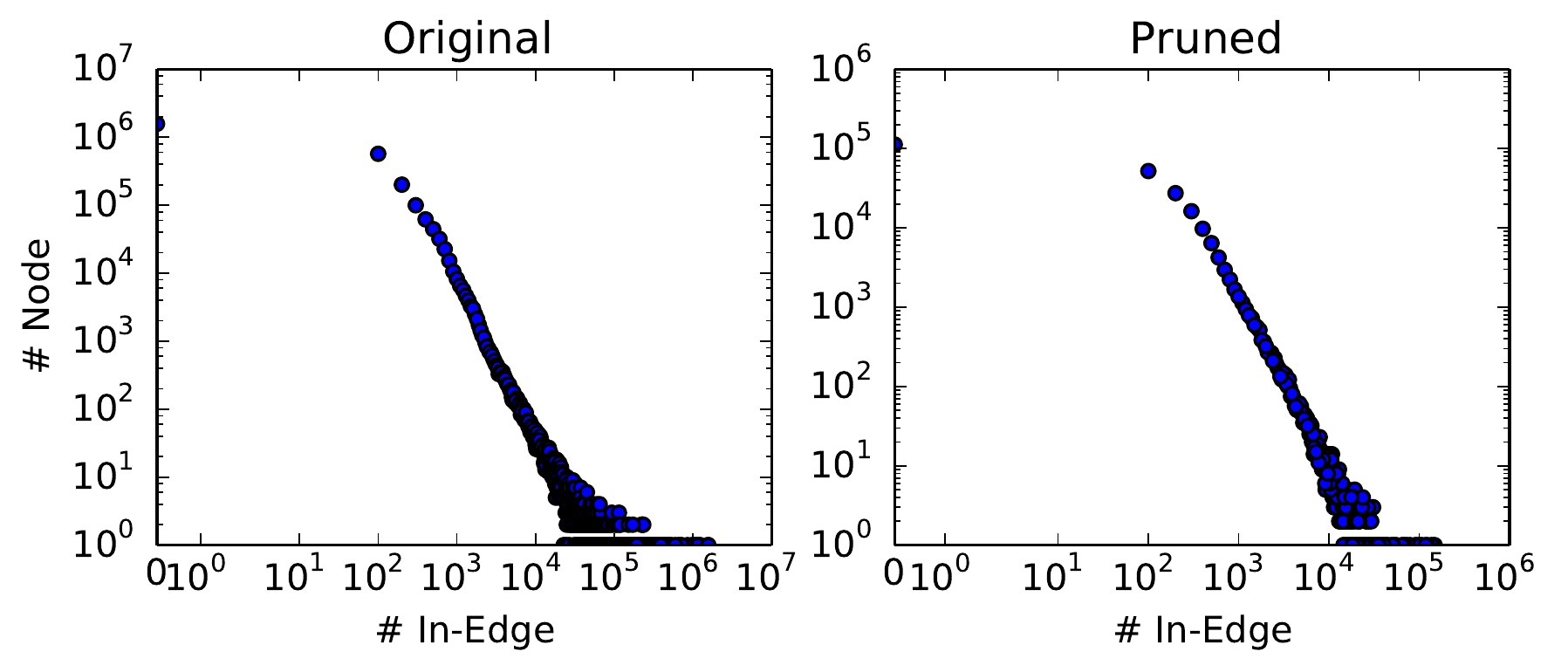}
\caption{In-edge distributions of the original network (on the left) and the
pruned network (on the right).}
\label{fig:inedge-distribution}
\end{figure}

\begin{figure}[t]
\centering
  \includegraphics[width=.5\linewidth,keepaspectratio]{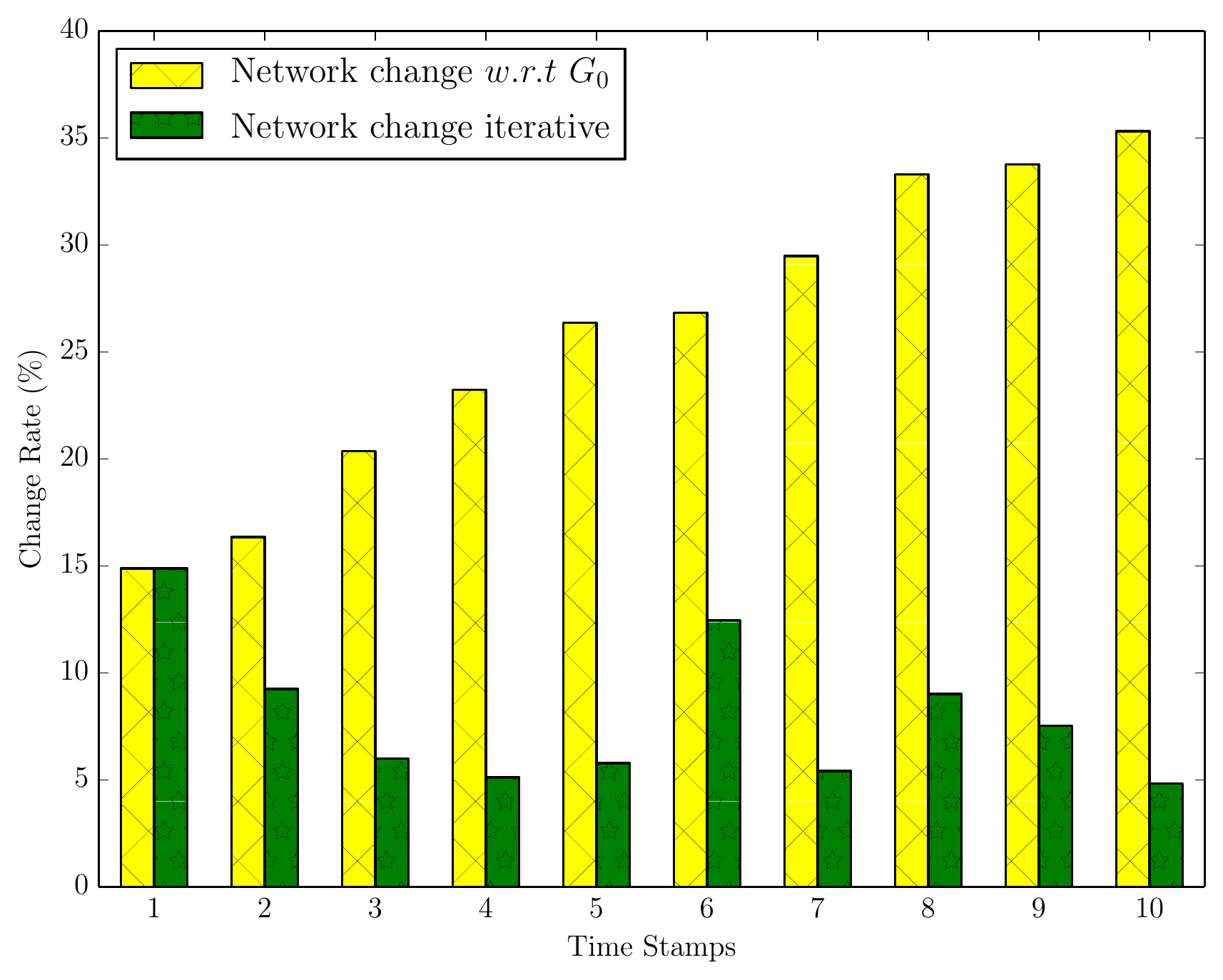}
  \caption{Change rate of the network over each iteration w.r.t the previous one and 
  w.r.t. the original one.}
  \label{fig:networkChange}
\end{figure}

Figure~\ref{fig:inedge-distribution} shows the in-edge distribution of the
original and the pruned network. Both follow a power-law distribution. Impact
of the  pruning process on the network structure seems to be minimal and has
not created any anomalies in the analysis. We also pruned the tweet list
according to the same top $250K$ influential users, which reduced the total
size of the tweet sets to $200M$.
Figure~\ref{fig:networkChange} shows how much the network has
changed over each iteration with respect to the previous snapshot ($\frac{|E_t
\backslash E_{t-1}|}{|E_{t-1}|}$) and with respect to the original one
($\frac{|E_t \backslash E_0|}{|E_0|}$). Here, change w.r.t. previous snapshots
is defined in order to have an insight about the experimental data and it
cannot be compared with the experimental results of the any probing strategy.
It represents the case where exact snapshots of the network exist locally,
which is not the case in a real-world scenario. In a probing scenario where
the exact network is not available, network error is expected to increase, as
we are continuously building on top of the previous partial network which also
contains some amount of error. Therefore, iterative change w.r.t. original
network better matches a real-world scenario.

\subsection{Evaluation of Dynamic Network Fetching}
\noindent
We have implemented several algorithms to compare the performance of the
proposed techniques. The details of the algorithms used are given as follows:
\smallskip
\\
\textbf{NoProbe} and \textbf{Random Probing}. These are two baseline
algorithms. \textit{NoProbe} algorithm assumes that the network does not
change over time and uses the fully observed network at time $t=0$ for all time
points without performing any probing. It represents the worst case scenario
for dynamic network fetching. The second baseline algorithm is
\textit{Random Probing} algorithm which randomly chooses $k$ users to probe
with uniform probability. In the experiments, this baseline
method is run 10 times and the average values of these runs are used in the
evaluation.%
\smallskip
\\
\textbf{Indegree Probing.} This is our third baseline algorithm 
that uses a very similar idea to our proposed technique from Eq.~\ref{Score}. This 
baseline method utilizes the same formula with one change, instead of using PageRank
values it uses the indegree values of the users 
($Score(v) = (1-\theta)Deg'_{v}(t-1) + \theta\,\sigma_{IP^{Deg}_{v}}$).%
\smallskip
\\
\textbf{MaxG.} As described in \cite{dynamicinfmax}, users are probed with a
probability proportional to the ``performance gap'', which is defined as the
predicted difference between the results of the approximate solution and the
real solution. Briefly, the method incrementally probes users which will bring
the largest difference in the results. It assumes that the influence
of a specific user is related to the output of the \emph{degree discount
heuristic}. Although their influence determination function is different than
ours, we use the MaxG algorithm for performance evaluation of our proposed
algorithms.%
\smallskip
\\
\textbf{Priority Probing.} As described in \cite{bahmani2012pagerank}, this 
algorithm chooses users to probe according to a value proportional to their 
priorities. Priority of a node is defined as the value of its PageRank score. 
For every iteration of the method, if a node is not probed, the current 
PageRank value is added to its priority and if the node is probed, its 
priority is reset to 0.%
\smallskip
\\
\textbf{Change Probing.} This is our first proposed method, which chooses $k$
users to probe with value proportional to their scores, as computed by
Eq.~\ref{Score}. The network is then constructed via
Alg.~\ref{Alg_Network_Fetch}.%
\smallskip
\\
\textbf{RRCh Probing.} This is our second proposed method,
which chooses $\beta\cdot k$ users to probe with Change Probing and
$(1-\beta)\cdot k$ users with Round-Robin Probing. When $\theta=0$ in
Eq.~\ref{Score} for the Change Probing part, the method becomes similar
to~\cite{bahmani2012pagerank}. The difference is that Priority Probing
increases the probe possibility of a node by its PageRank value in every step
if it is not probed, so that at some point the probe possibility becomes $1$.%

We evaluate performance by comparing the quality of the influential users
found by each approach with that of the ideal case. For this purpose, we use
two different evaluation measures:
 \begin{itemize}
   \item Jaccard similarity between the correct and estimated top-$k$ most
   influential users lists.
   \item The mean squared error (Eq.~\ref{MSE}) of the PageRank scores. 
   The reported values with respect to the probing capacities of MSE are the 
   average values of all 11 snapshots. The values with respect to time are
   the average values of different probing capacities. Additionally, standard 
   deviations of the values are also reported in the discussions.
 \end{itemize}

\begin{equation}\label{MSE}
  MSE = \sqrt{\frac{1}{|V_t \cap V_{t}^{'}|} \sum_{ \forall v \in V_{t}^{'} \cap V_t}{(PR^{'}_{t}(v) - PR_{t}(v))^2}}
\end{equation}

\subsection{Evaluation of Dynamic Tweet Fetching}
\noindent
We evaluate the performance of the proposed tweet fetching technique with two 
baselines algorithms, namely \textit{NoProbe} and \textit{Random Probing}. The 
details of these baselines are given below:
\smallskip
\\
\textbf{NoProbe.} This algorithm assumes that the tweet set does not change
over time and use the fully observed tweet set at time $t=0$ for all time
points without any probing. This method represents the worst case scenario for
the dynamic tweet fetching problem.
\smallskip
\\
\textbf{Random Probing.} This algorithm randomly chooses $k$ users to collect 
tweets with uniform probability at each time step.
\smallskip
\\
\textbf{RRCh Probing.} This is the
algorithm we proposed, which greedily chooses $k$ users to collect tweets with
value proportional to their scores describe in Eq.~\ref{Score}. Differently
from the network fetching method, scores are calculated by using $WPR_{v}^{j}$
for the topic $C_j$, instead of $PR_{v}$.

\subsection{Experimental Results and Discussion}
\noindent
This section compares and discusses the performance of the proposed network
and tweet probing methods with the state-of-the-art and baseline methods
using experiments executed on real datasets. We also provide an empirical
interpretation of the calculated topic-based influence scores.

\subsubsection{Experimental Setup}
\noindent
As indicated by Eqs.~\ref{limitationRelFormula}~and~
\ref{limitationTextFormula}, given the resource limits permitted by the 
service providers, one cannot probe a significant portion of the network. We 
have executed our experiments with different probing capacities and used $0.001
\%$, $0.01\%$, $0.1\%$ and $1\%$ of the network as the size of the probe set.
For the analysis of the effect of the $\theta$ parameter used in Change
Probing, we set: 
\begin{enumerate*}[label=\itshape\alph*\upshape)] 
\item $\theta=0$,
meaning PageRank proportional scores are used;
\item $\theta=0.5$,
meaning equally weighted PageRank and influence past scores are used;
\item $\theta=1$,
meaning only influence past scores are used
\end{enumerate*}. 
For the RRCh algorithm we tested the ratio parameter $\beta$
with three values, which control the fraction of vertices proved via random
selection: $0.4$, $0.6$, and $0.8$.

\subsubsection{Change Probing Performance w.r.t. $\theta$}
\noindent
Figure~\ref{fig:chPerformance} depicts the performance of Change Probing
algorithm for the average Jaccard similarity and MSE measures. As expected, Change Probing
algorithm significantly outperforms \textit{NoProbe} algorithm. For the optimization of
the $\theta$ parameter, we test Change Probing algorithm under three different
$\theta$ configurations:

\begin{figure}[t]
\centering
\begin{subfigure}{.5\textwidth}
  \centering
  \includegraphics[width=.45\linewidth,keepaspectratio]{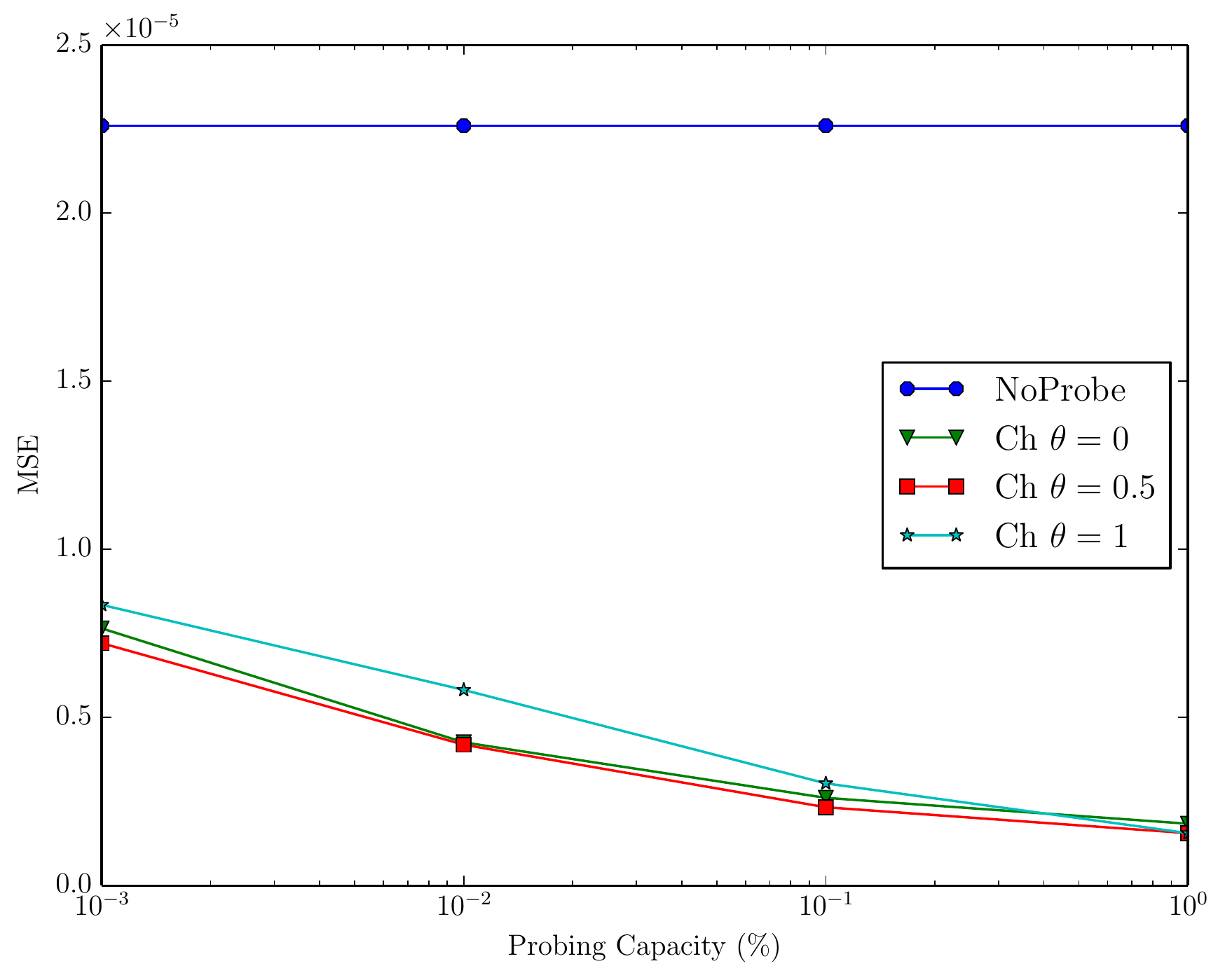}
  \caption{Average MSE for all snapshots.}
  \label{fig:mseCh}
\end{subfigure} \\
\begin{subfigure}{.5\textwidth}
  \centering
  \includegraphics[width=.75\linewidth,keepaspectratio]{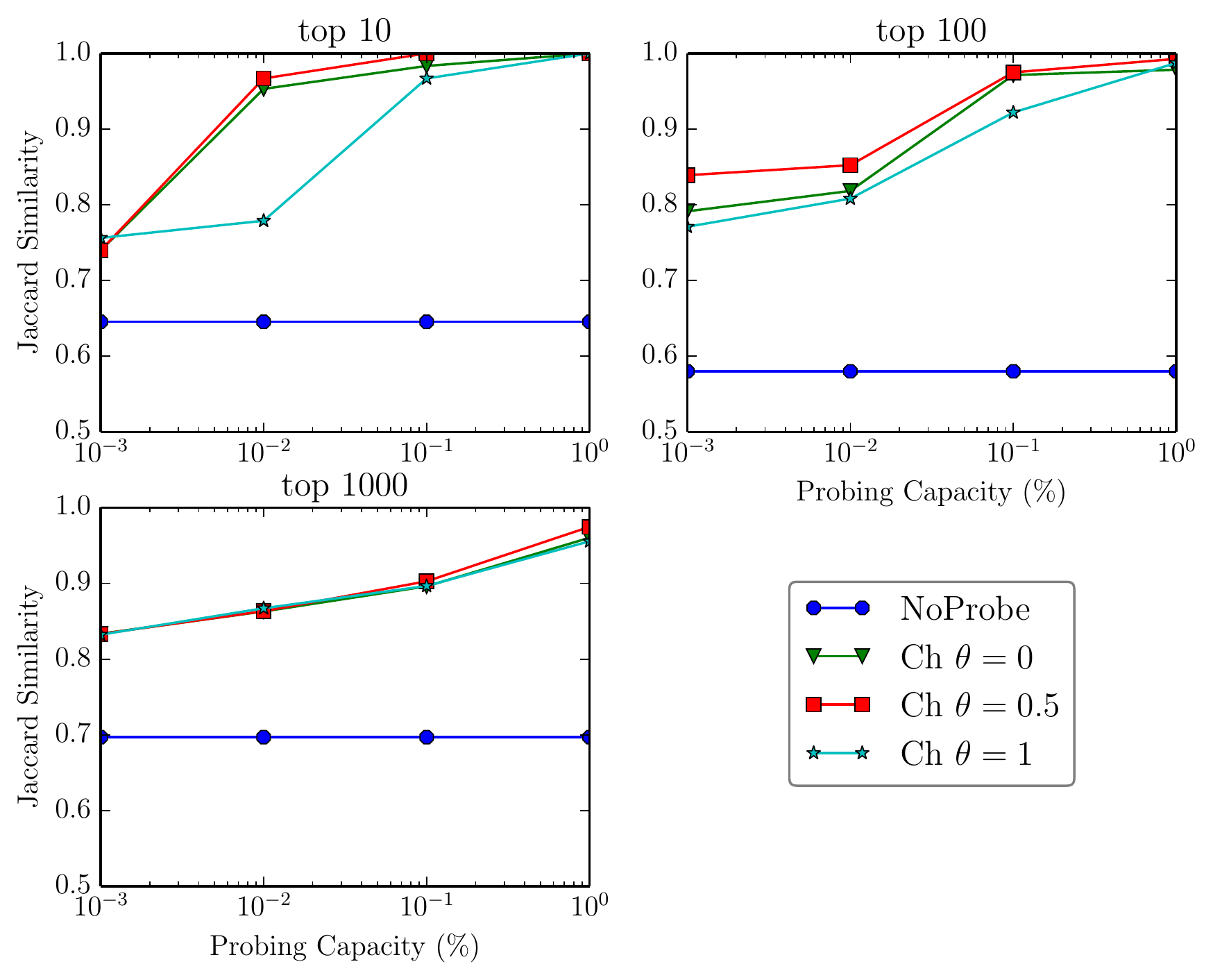}
  \caption{Average Jaccard similarity for all snapshots.}
  \label{fig:jacCh}
\end{subfigure} 
\caption{Performance of Change Probing w.r.t. $\theta$.}
\label{fig:chPerformance}
\end{figure}

\begin{itemize}
\item Using the MSE measure, $\theta=0.5$ setting performs $8\%$ better than
$\theta=0$ setting and $19\%$ better than $\theta=1$ setting. Overall, it
performs $83\%$ better than \textit{NoProbe}.

\item Using the Jaccard distance measure, $\theta=0.5$ setting is $3\%$ better than
$\theta=0$ setting and $5\%$ better than $\theta=1$ setting. In the overall
case, $\theta=0.5$ outperforms \textit{NoProbe} by $43\%$. We also note that as the
probing capacity increases, performance of the Change Probing algorithm
becomes less dependent on the setting of $\theta$. 
\end{itemize}

\begin{figure}[t]
\centering
\begin{subfigure}{.5\textwidth}
  \centering
  \includegraphics[width=.75\linewidth,keepaspectratio]{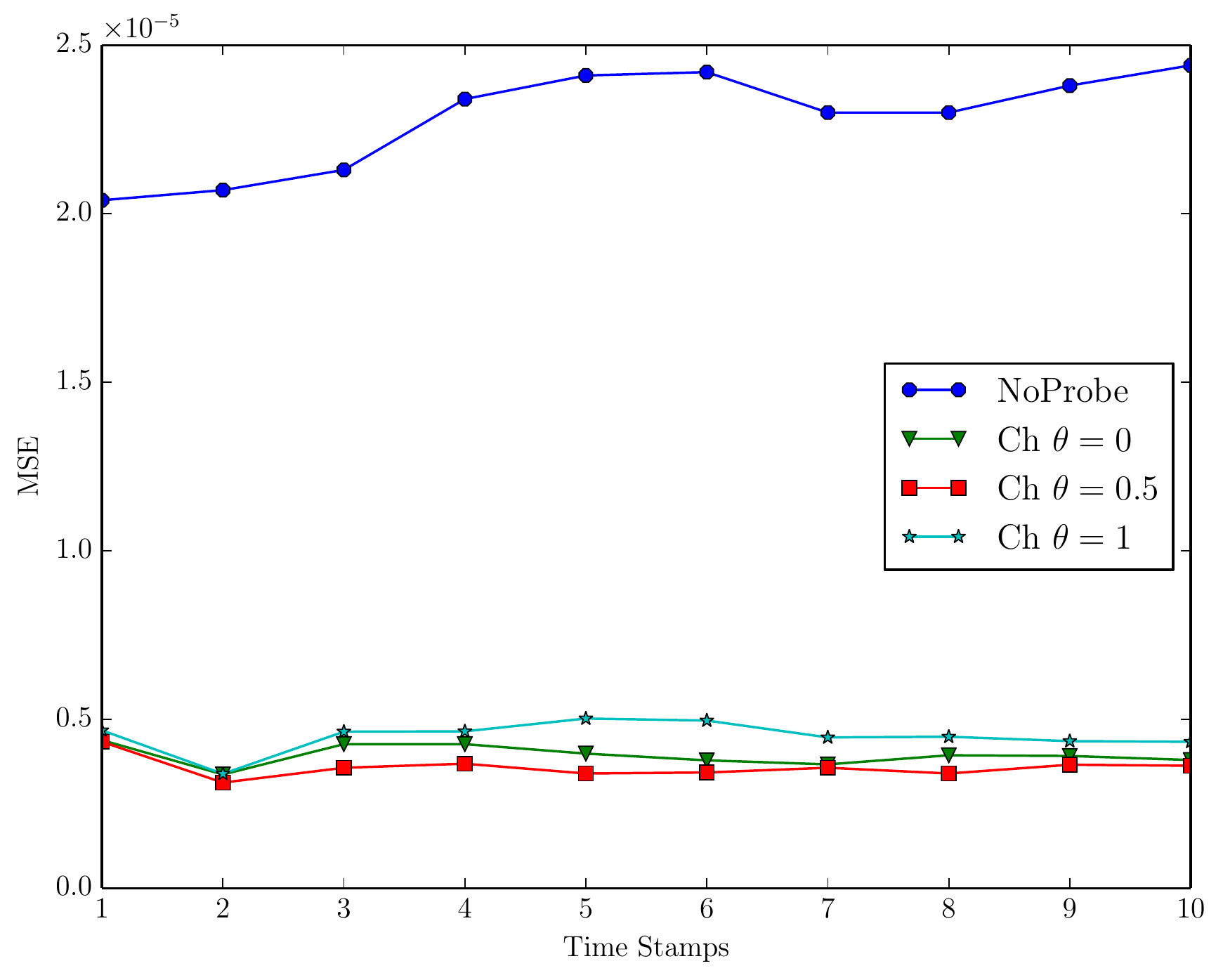}
  \caption{Average MSE for all probing capacities.}
  \label{fig:mseTimeCh}
\end{subfigure} \\
\begin{subfigure}{.5\textwidth}
  \centering
  \includegraphics[width=.75\linewidth,keepaspectratio]{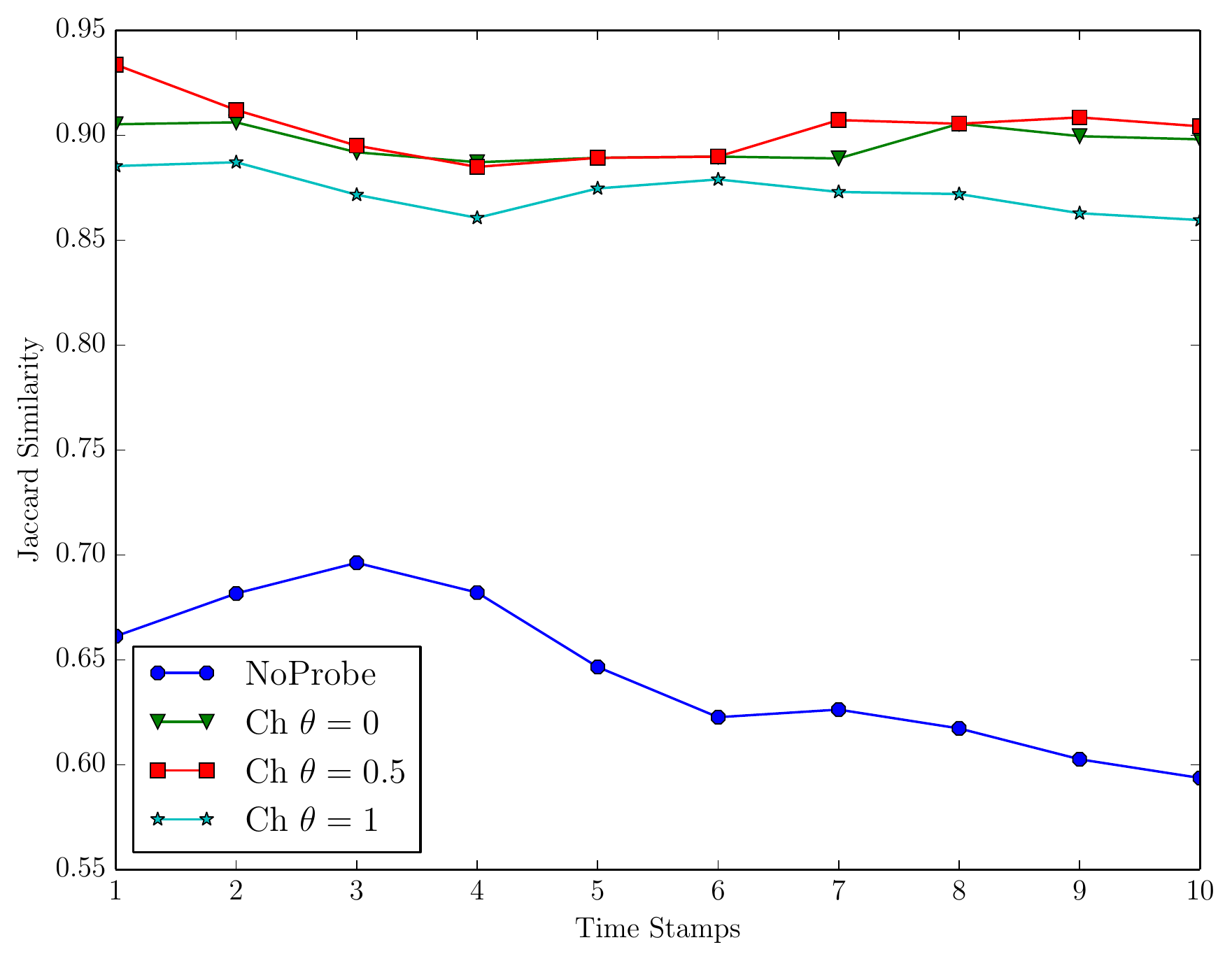}
  \caption{Average Jaccard similarity for all probing capacities.}
  \label{fig:jacTimeMSECh}
\end{subfigure} 
\caption{Performance of Change Probing as a function of time.}
\label{fig:chPerformanceInTime}
\end{figure}

We also illustrate the change in error as the network evolves, in order to see
how the performance of different algorithms are affected as the seed network
data ages. Figures~\ref{fig:mseTimeCh}~and~\ref{fig:jacTimeMSECh}\footnote{
Jaccard similarity reports the
average values of all three probing capacity settings.} show the performance
of Change Probing as a function of time for the mean squared error (MSE) and
Jaccard similarity measures, respectively. We observe that \textit{NoProbe}
has an increasing error as time passes. Change Probing gives a more robust and
stable performance with respect to time. As the number of past influence
points increases, the algorithm can estimate the influence variability of the
users more accurately, which compensates the deteriorating effect of aging of
the baseline network data. Since $\theta=0.5$ outperforms the other cases, we
use $\theta=0.5$ configuration in the subsequent experiments with other
algorithms. We also note that y-axis contains relatively small values because
the PageRank values are normalized. We have assumed \textit{NoProbe} algorithm
as the reference point for normalization.

\subsubsection{RRCh Probing Performance w.r.t. $\beta$}
\begin{figure}[t]
\centering
\begin{subfigure}{.5\textwidth}
  \centering
  \includegraphics[width=.45\linewidth,keepaspectratio]{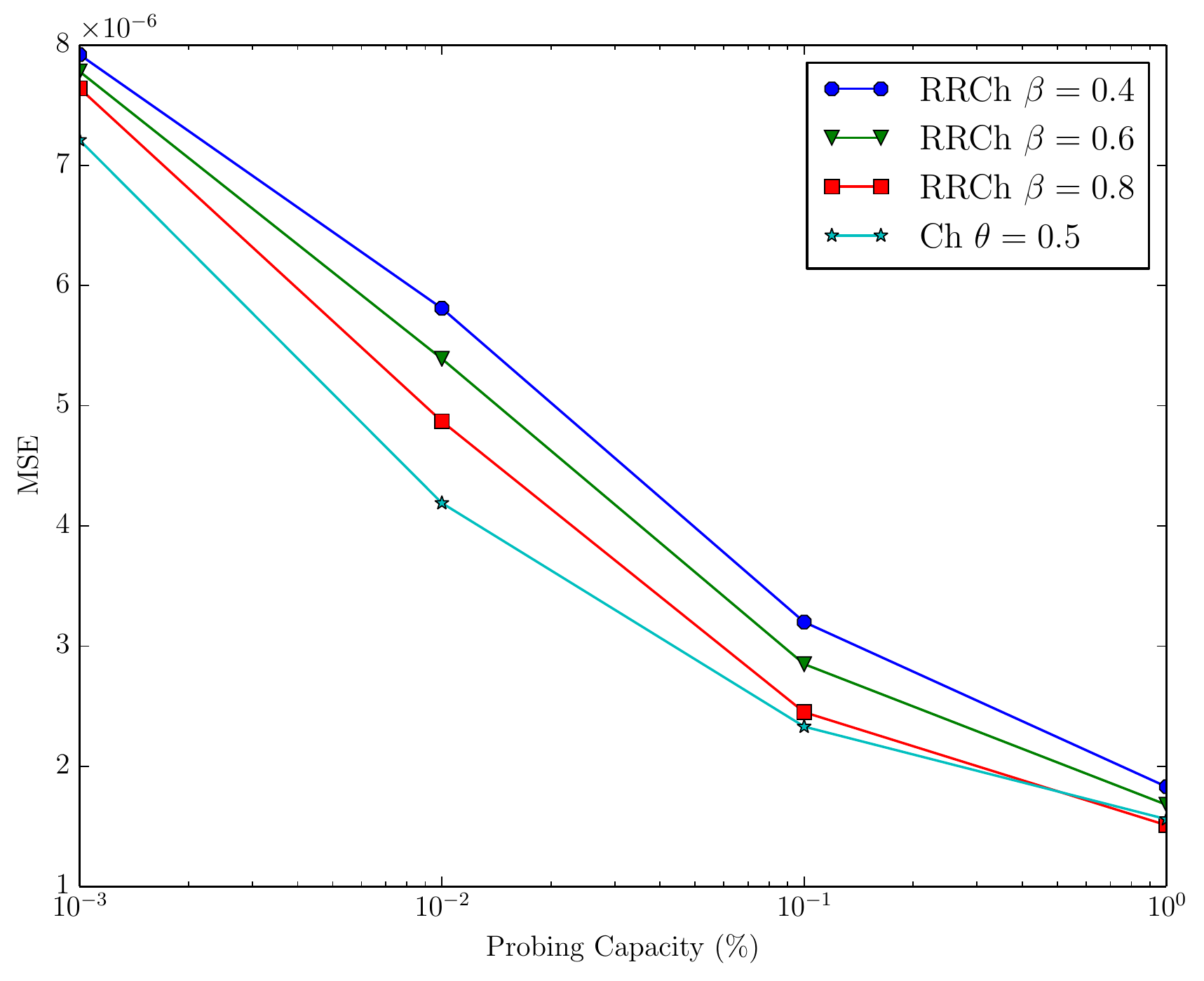}
  \caption{Average MSE for all snapshots.}
  \label{fig:mseRrch}
\end{subfigure} \\
\begin{subfigure}{.5\textwidth}
  \centering
  \includegraphics[width=.75\linewidth,keepaspectratio]{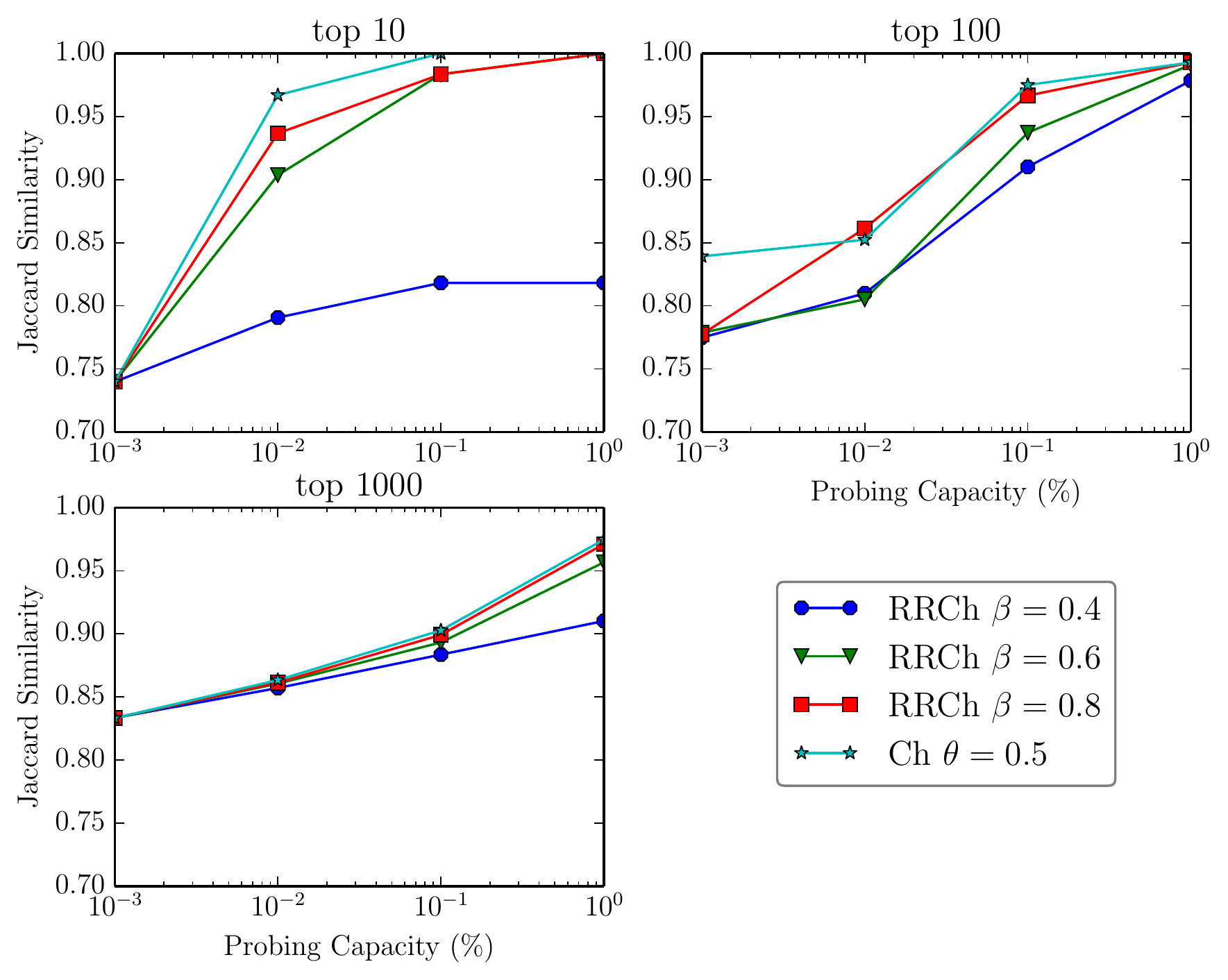}
 \caption{Average Jaccard similarity for all snapshots.}
 \label{fig:jacRrch}
\end{subfigure}
\caption{Performance of RRCh w.r.t. $\beta$.}
\label{fig:rrchPerformance}
\end{figure}

\noindent
Figure~\ref{fig:rrchPerformance} shows the performance results for the
Round-Robin Change (RRCh) Probing algorithm under different round-robin
ratios. We use the Change Probing algorithm (with $\theta=0.5$ setting) as the
baseline reference point.

\begin{figure}[t]
\centering
\begin{subfigure}{.5\textwidth}
  \centering
  \includegraphics[width=.75\linewidth,keepaspectratio]{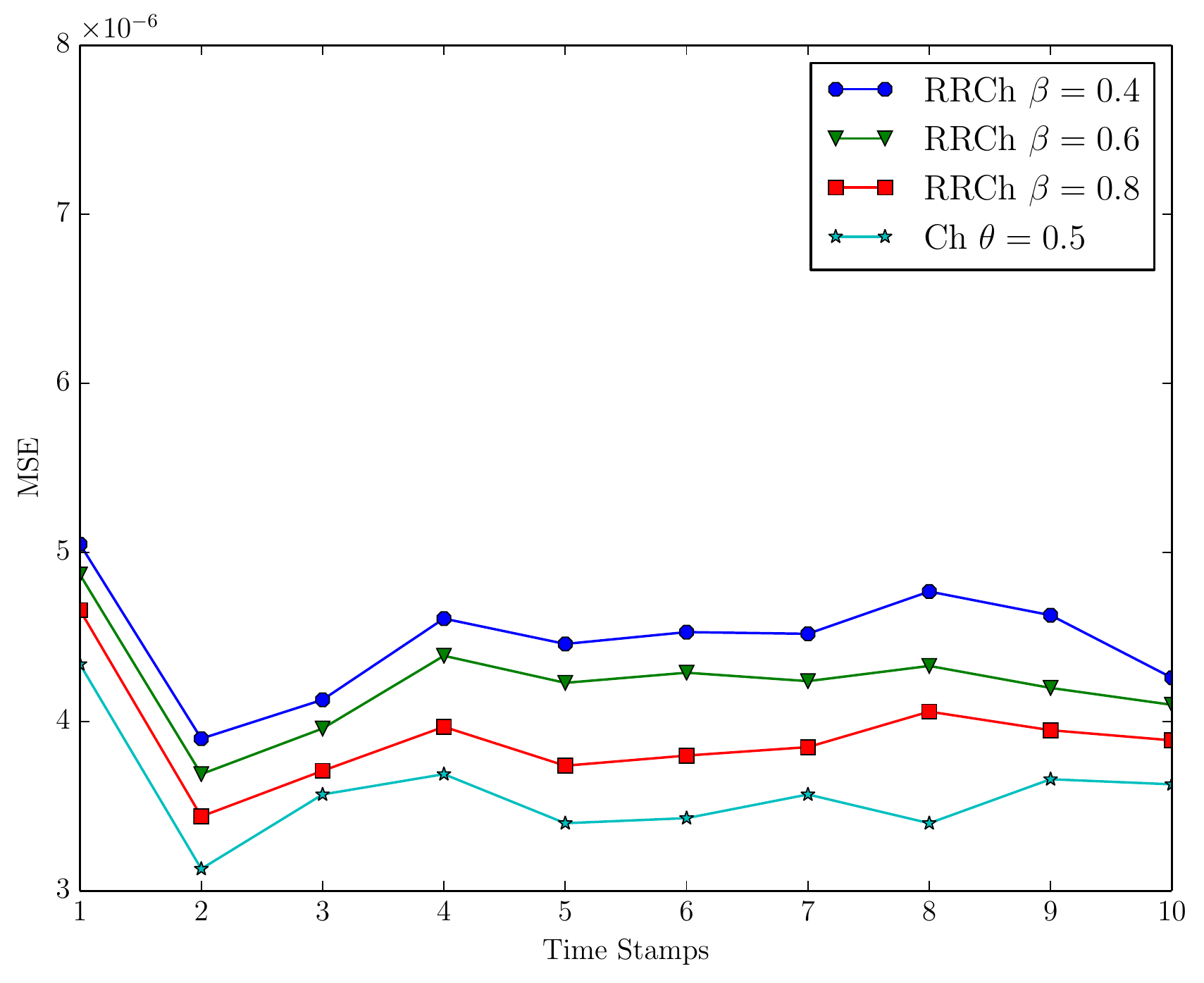}
  \caption{Average MSE for all probing capacities.}
  \label{fig:mseTimeRrch}
\end{subfigure} \\
\begin{subfigure}{.5\textwidth}
  \centering
  \includegraphics[width=.75\linewidth,keepaspectratio]{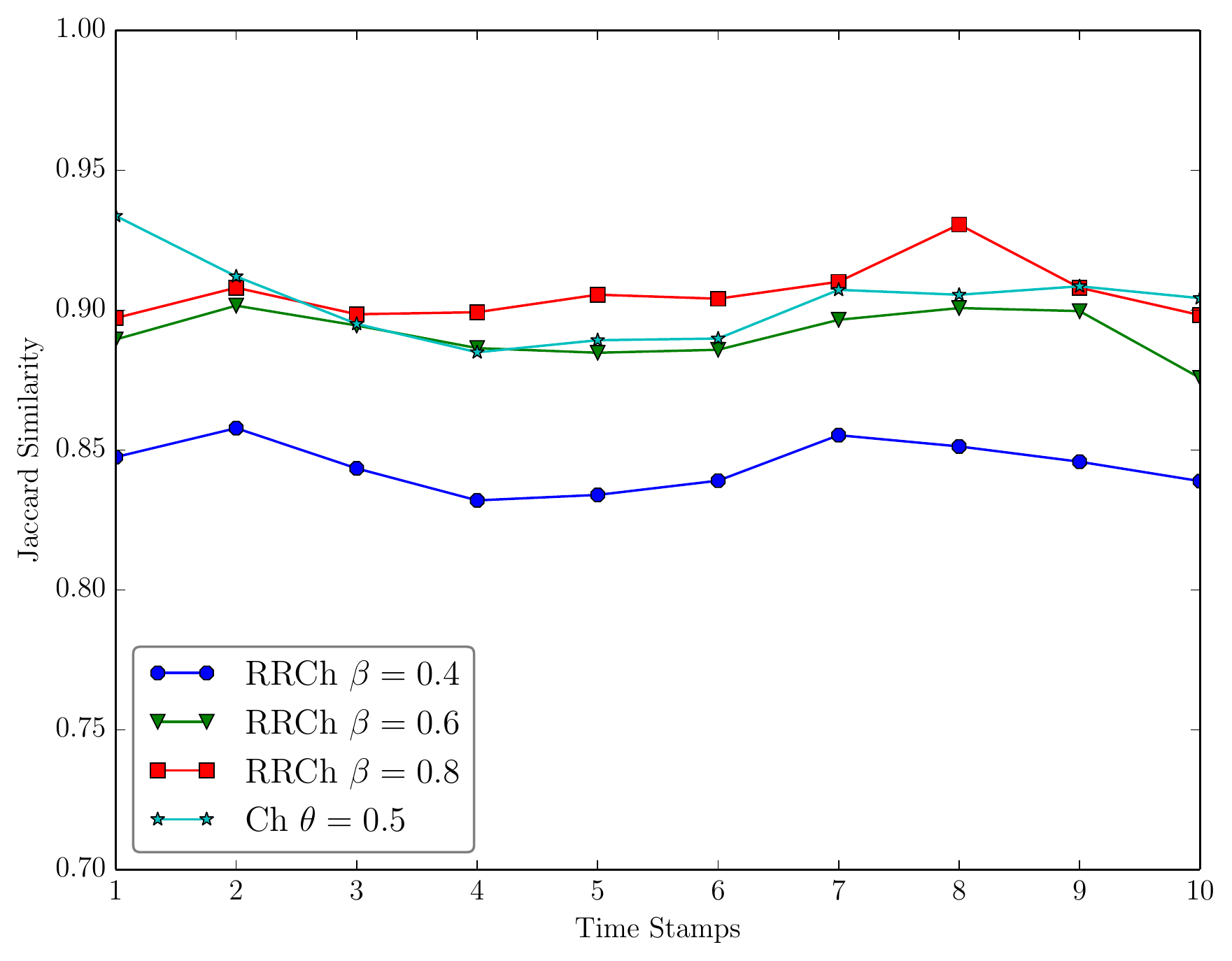}
  \caption{Average Jaccard similarity for all probing capacities.}
  \label{fig:jacTimeMSERrch}
\end{subfigure}
\caption{Performance of RRCh as a function of time.}
\label{fig:rrchPerformanceinTime}
\end{figure}

We observe that the RRCh algorithm performs poorly for small probing
capacities, such as $0.001\%$ and $0.01\%$. Randomness impacts the performance
more with smaller number of probed users, since we are not able to probe the
influential users with great influential power, thus lowering the performance.
For MSE, $\beta=0.8$ configuration performs $7\%$ better than $\beta=0.6$ and
$12\%$ better than $\beta=0.4$. For the Jaccard similarity measure, it is
$2\%$ better than $\beta=0.6$ and $7\%$ better than $\beta=0.4$. Although, it
performs worse than Change Probing in the short term, it reaches the
performance of Change Probing in the long term, as show in in
Figures~\ref{fig:mseTimeRrch}~and~\ref{fig:jacTimeMSERrch}. Moreover, it
guarantees the probing of every node within a time frame, preventing the
system to focus on only a limited section of the network and missing other
regional changes that might accumulate and start to affect the network in the
global sense. We would have seen this phenomenon more explicitly if the number
of snapshots were larger, which was the case in~\cite{dynamicinfmax}. The
results are slightly better when the ratio is set to $\beta=0.8$. Therefore,
we choose to use this algorithm (with $\theta=0.5$ and $\beta=0.8$
configurations) instead of Change Probing for the comparison with others in
the following sections.

\begin{figure}[t]
\centering
  \includegraphics[width=.5\linewidth,keepaspectratio]{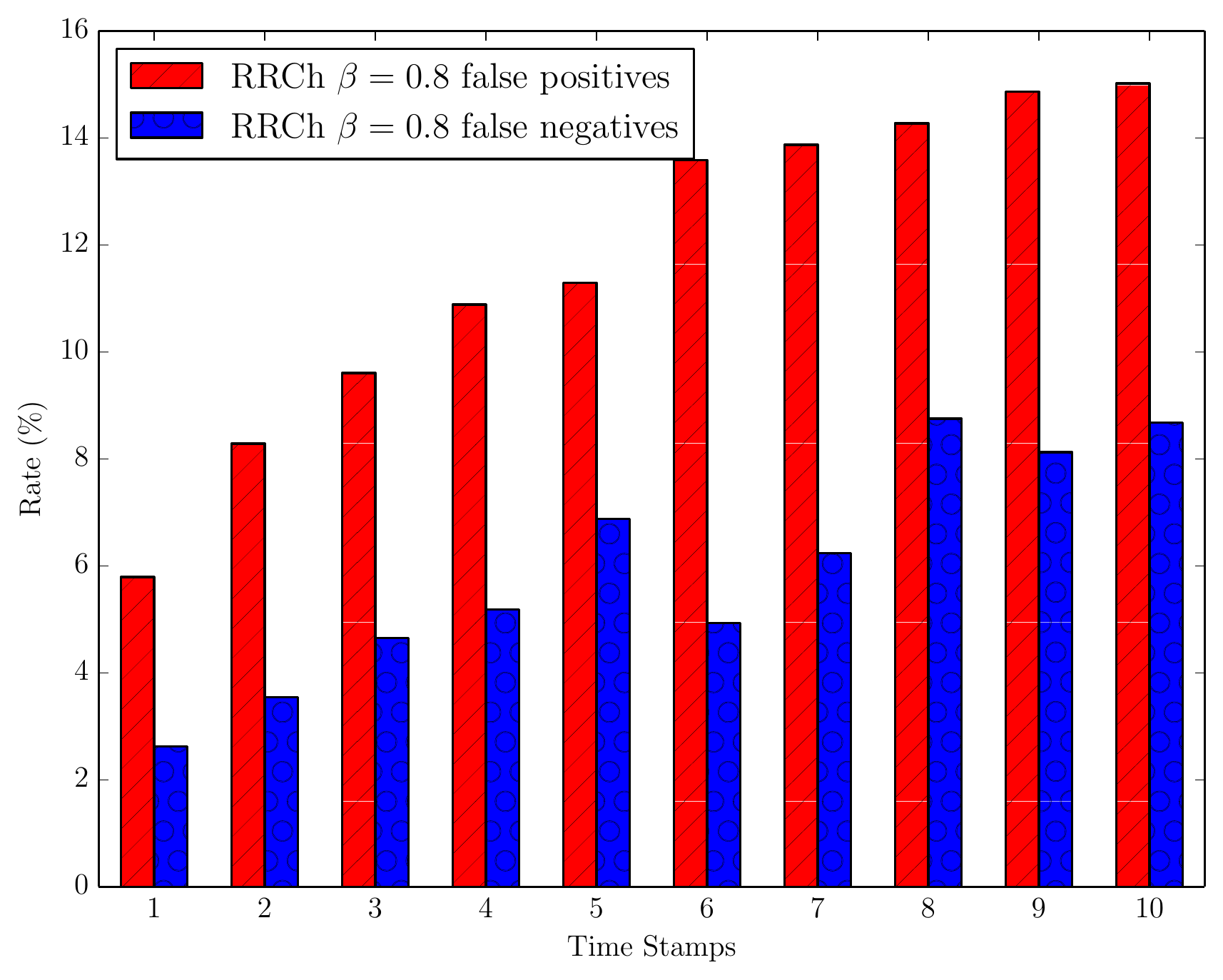}
  \caption{False positives and false negatives rates for every snapshot in time.}
  \label{fig:posNegRates}
\end{figure}

Figure \ref{fig:posNegRates} shows both the percentages of edges that were not 
present in the the true network but were assumed to be present by the
algorithm (false positives) and the percentages of edges that were present in
the true network but were not captured by the algorithm (false negatives). The
findings indicate that the proposed technique is doing a good job at capturing
the structure of the network by having on average $12\%$ false positives and $6\%$
false negatives rates for all snapshots.

\subsubsection{Comparison with the State-of-the-Art}
\begin{figure}[t]
\centering
\begin{subfigure}{.5\textwidth}
  \centering
  \includegraphics[width=.45\linewidth,keepaspectratio]{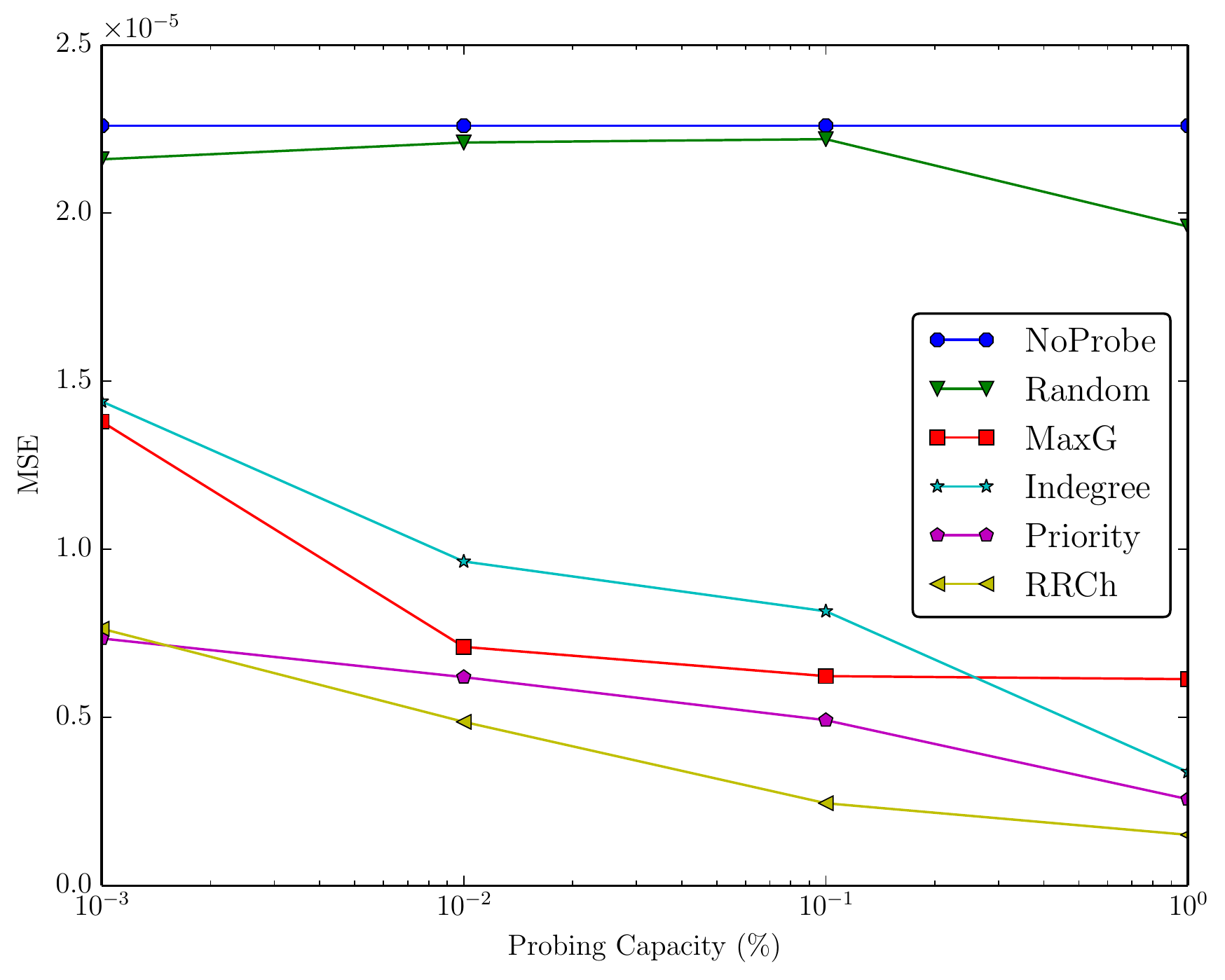}
  \caption{Average MSE for all snapshots.}
  \label{fig:mseAll}
\end{subfigure} \\
\begin{subfigure}{.5\textwidth}
  \centering
  \includegraphics[width=.75\linewidth,keepaspectratio]{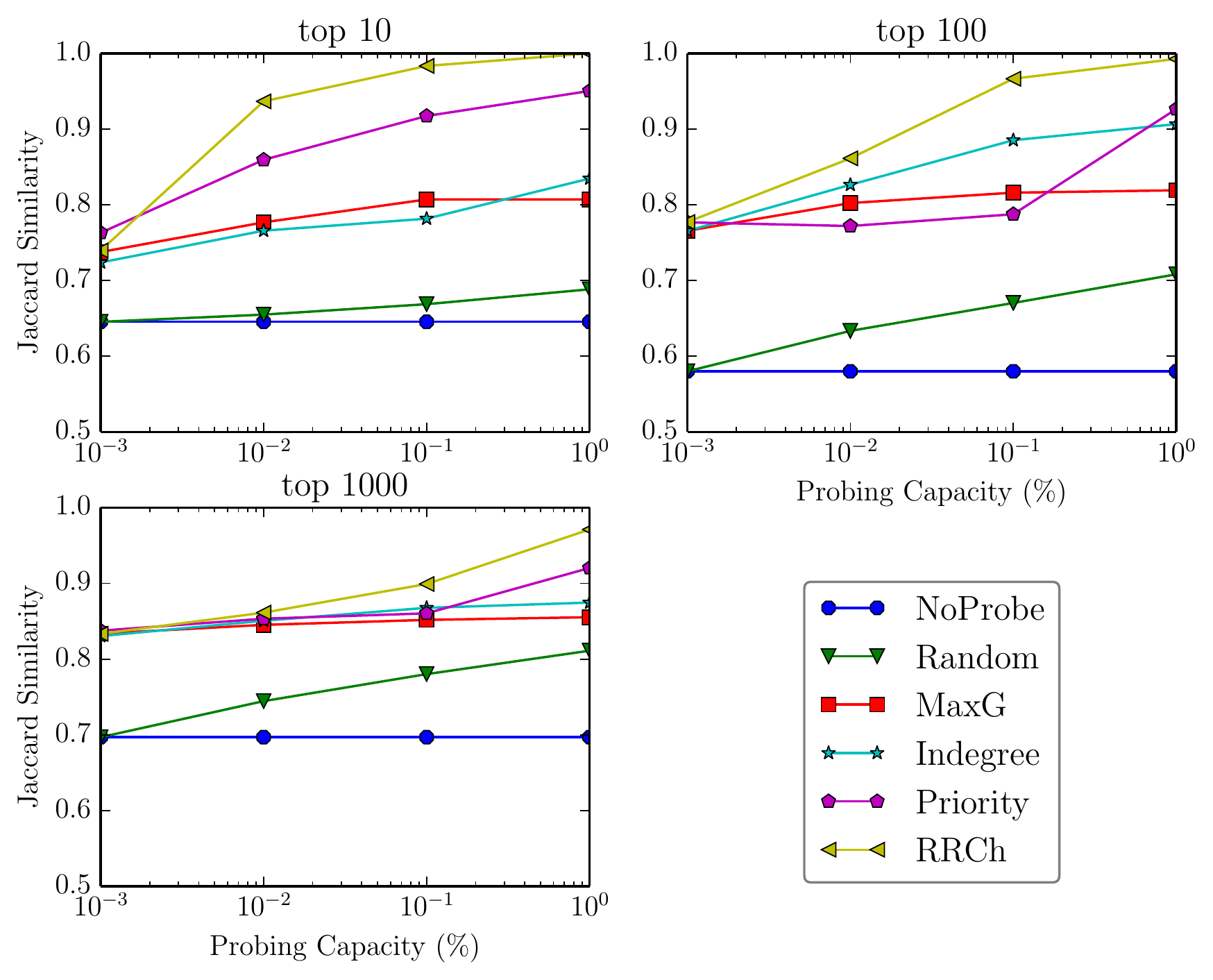}
  \caption{Average Jaccard similarity for all snapshots.}
  \label{fig:jacAll}
\end{subfigure}
\caption{Comparison of the probing strategies.}
\label{fig:allPerformance}
\end{figure}

\noindent
Figure~\ref{fig:allPerformance} compares the performance of RRCh method
(with $\theta=0.5$ and $\beta=0.8$ settings) against the baselines and the
state-of-the-art methods from the literature. RRCh achieves better
results for all performance measures used for comparison in our paper. It
reduces MSE by $21\%$ (see Figure~\ref{fig:mseAll}) when compared to Priority
Probing, $41\%$ when compared to Indegree Probing and $49\%$ when compared to 
the MaxG method. Priority Probing suffers
especially for low probing capacities, since the priority of a user is set
to $0$ after probing. A probed user can regain its priority very late in the
process, which prevents it to track quick changes in the scores of the highly
influential users. Therefore, after probing an important user in terms of
influence, that user is not being probed for some time, even if the influence
of the user is changing very fast. RRCh always probes $\beta$ portion of
the users according to their influence impact and change over time, so that
the important users are in the probe set at each step.


\begin{figure}[t]
\centering
\begin{subfigure}{.5\textwidth}
  \centering
  \includegraphics[width=.75\linewidth,keepaspectratio]{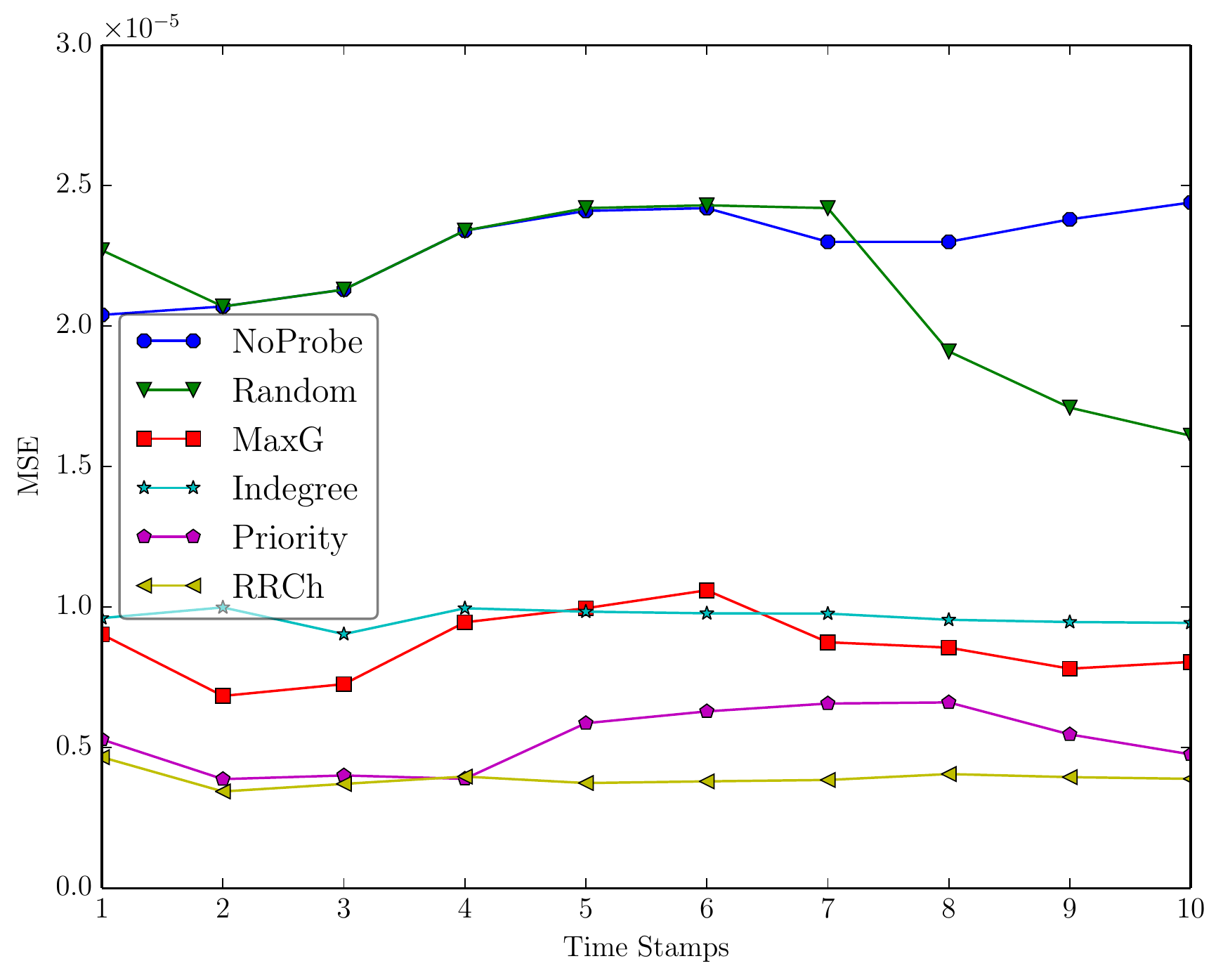}
  \caption{Average MSE for all probing capacities.}
  \label{fig:mseTimeAll}
\end{subfigure} \\
\begin{subfigure}{.5\textwidth}
  \centering
  \includegraphics[width=.75\linewidth,keepaspectratio]{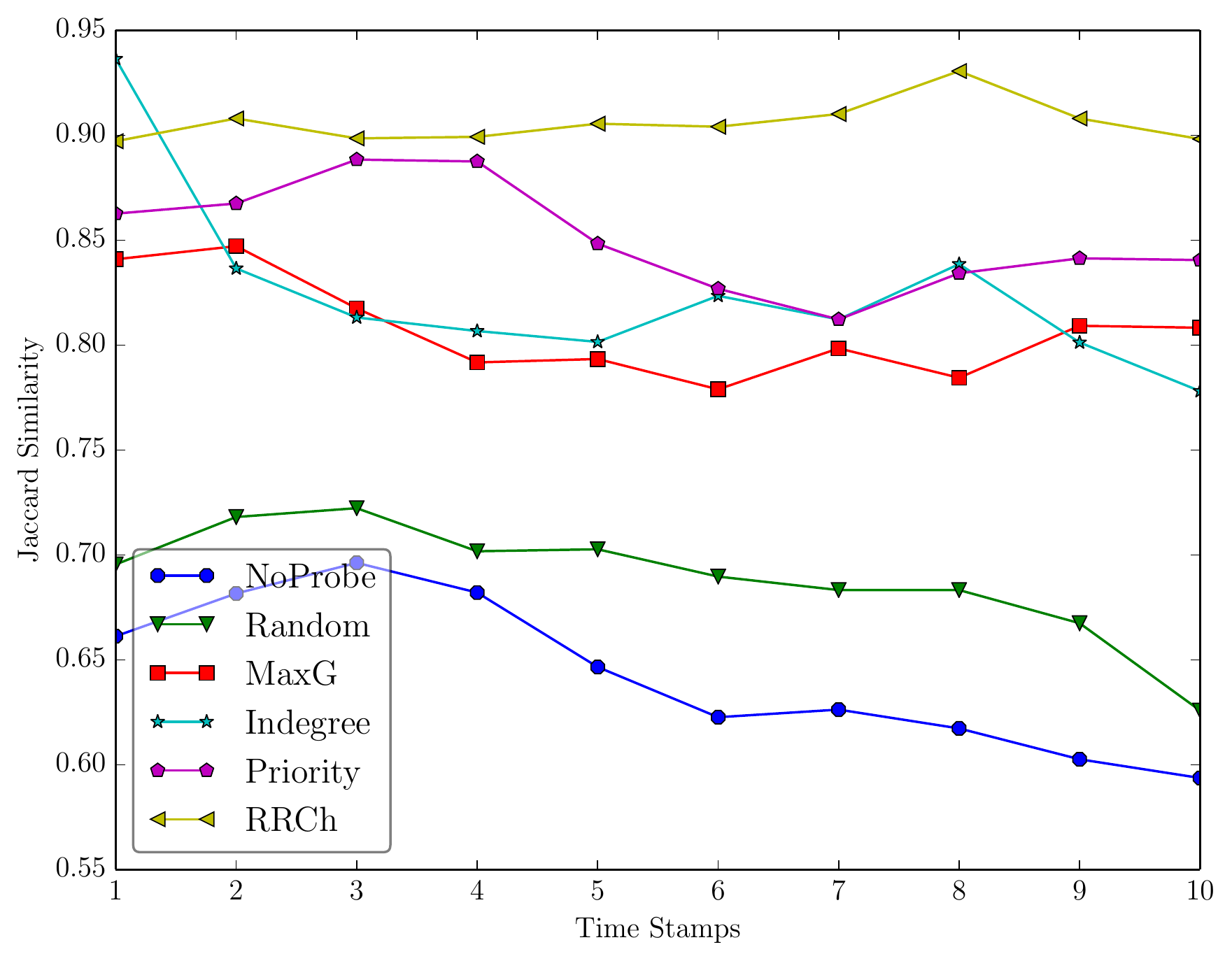}
  \caption{Average Jaccard similarity for all probing capacities.}
  \label{fig:jacTimeMSEAll}
\end{subfigure}
\caption{Comparison of the Probing strategies with respect to time.}
\label{fig:allPerformanceinTime}
\end{figure}

Overall, our proposed method gives $80\%$ higher performance than the NoProbe
and Random Probing algorithms for the MSE measure. As seen in Figure~\ref{fig:jacAll}, RRCh
shows better results for the top-k set similarities as well. It is $5\%$
better than Priority Probing, $7\%$ better than Indegree Probing and $11\%$ better than MaxG method on average. RR
Change performs $35\%$ better against baselines when Jaccard similarity is
considered. Since it also considers the change in the influence over time, it
is also able to preserve its accuracy while the performance of other methods
degrade over time (see Figures~\ref{fig:mseTimeAll}~and~\ref{fig:jacTimeMSEAll}).

\begin{figure}[t]
\centering
\includegraphics[width=.75\linewidth,keepaspectratio]{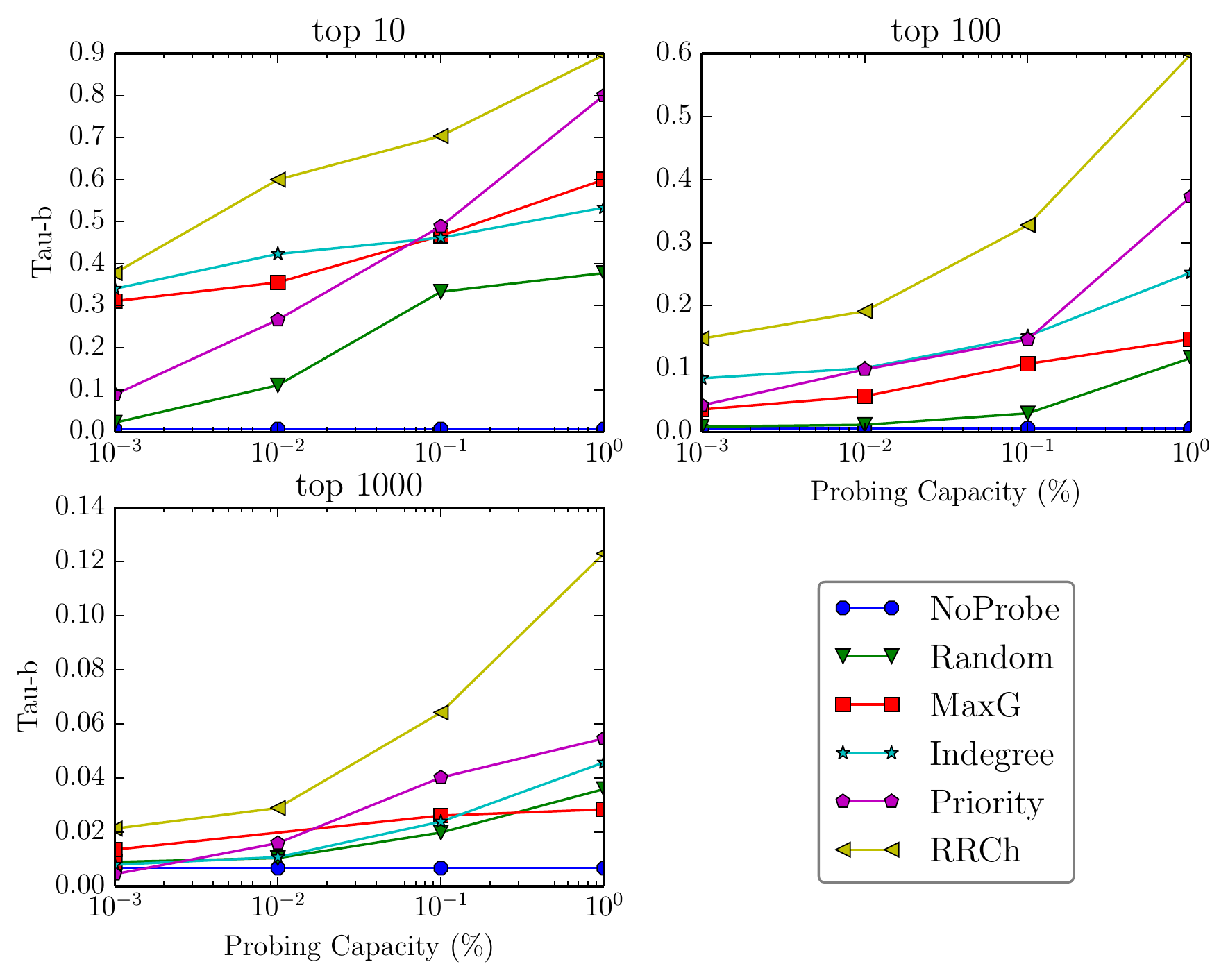}
\caption{Comparison of the probing strategies with respect to average Kendall Tau-b measure.}
\label{fig:allKendall}
\end{figure}

As mentioned before, in real-world scenarios one might not be
interested in the exact rank of the influential users but instead might select
top-k users and evaluate them by personal observation, because the ranking may
not be so accurate. Yet, we also compared the probing techniques against a
rank-aware similarity measure. Figure~\ref{fig:allKendall}  shows the
performance of alternative probing strategies based on the Kendall Tau-b
metric. The results are the average values from all of the snapshots. RRCh
gives $73\%$ higher performance than Random probing, $58\%$ higher than
Indegree Probing, $47\%$ higher than MaxG method and $40\%$ higher than
Priority Probing.

\subsubsection{Evaluation of the Network Inference Method} 
\begin{figure}[!ht]
\centering
\includegraphics[width=.5\linewidth]{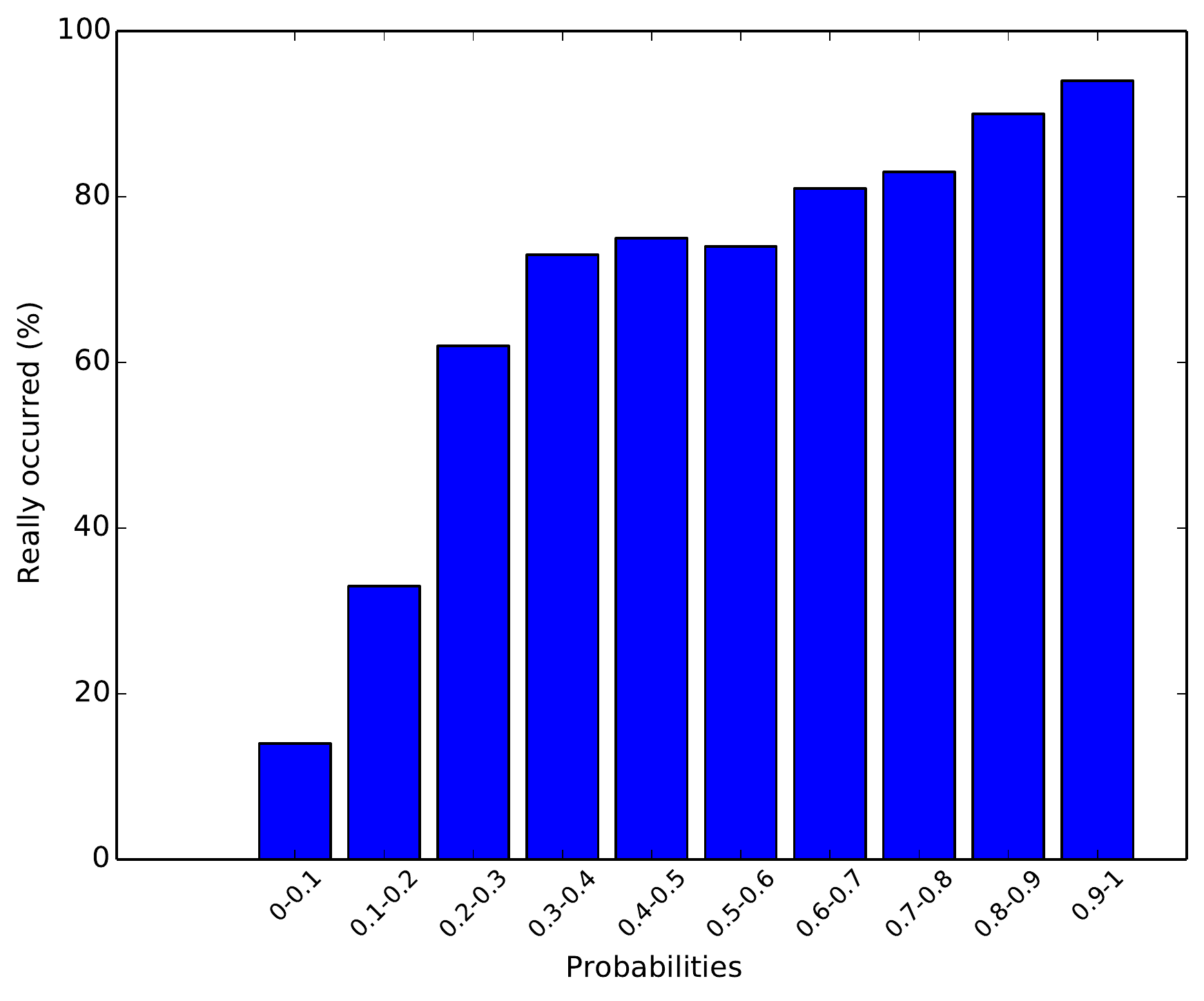}
\caption{Accuracy of the link prediction algorithm.}
\label{fig:predictionPercent}
\end{figure}

\noindent To assess
the prediction quality of the link prediction algorithm, we plotted the
histogram of the edges proposed by RA index that has really occurred in the
real network. This is shown in Figure~\ref{fig:predictionPercent}. The
histogram indicates the accuracy of the RA index used for network inference.
The edges that were determined by the prediction algorithm as more likely to
happen were found to be existent in the future network with a higher
probability. However, when we analyzed the incorrectly predicted edges, we
have observed that the algorithm predicts links between users who are unlikely
to follow each other in real life. For example, the algorithms predict an edge
between two pop stars since they have many common neighbors. However, they
would not follow each other because they are main competitors. Furthermore,
some of these users are not willing to follow anybody at all. This is the same
issue studied in~\cite{li2011casino}.  Link prediction algorithms typically do
not consider these facts in social networks. In addition to indexes which they
use to calculate similarities between users, they should also consider the
tendency of the users to make new connections. Therefore, we apply a filtering
process such that we only consider users who follow more than a threshold
number of users in order to determine users who are likely to follow somebody.
We add the predicted edges only to these selected users. As a result, we
improve the RRCh method by $3\%$ for MSE and $2\%$ for the set
similarities on average. Since the improvements are not
significant,  we omit the plots of those results for brevity. Here,
adaptation of more advanced (like mentioned in \ref{sec:dynamicdatafetching})
prediction algorithms could potentially increase the accuracy of this
technique. Moreover, the computational overhead of the link
prediction task is not significant due to the pruning process applied for the
experiments. The task takes less than a hour for one iteration. The time would
significantly increase for the size of original networks.

\subsubsection{Evaluation of the Topic Influence Estimation}\label{sec:EvalTopicInf}
\noindent
We evaluated the influence of users with respect to four different 
topics: 
\begin{enumerate*}[label=\itshape\alph*\upshape)] 
\item Politics, 
\item Sport, 
\item Health, and 
\item Cultural and Art Activities
\end{enumerate*}.
This section provides a qualitative discussion about the accounts which were 
found to be influential by the proposed methods. Table~\ref{influentialTable} 
shows the accuracy of topic relevance of the top-10 users found by the system 
for the specific topics.

\begin{table}[h!]
\centering
\begin{tabular}{|c|c|c|}
\hline
\textbf{Topics}      & \textbf{Topic Relevance}   & \textbf{Some selected accounts}\\ \hline
\textit{Politics}    & \textit{10 out of 10}      & \textit{\begin{tabular}[c]{@{}c@{}}RT\_Erdogan, kilicdarogluk, \\ 06melikgokcek\end{tabular}}    \\ \hline
\textit{Sport}       & \textit{8.5 out of 10}     & \textit{\begin{tabular}[c]{@{}c@{}}Fenerbahce, GalatasaraySK, \\ ntvspor\end{tabular}}   \\ \hline
\textit{Health}      & \textit{4 out of 10}       & \textit{\begin{tabular}[c]{@{}c@{}}saglikbakanligi, YYD\_tr, \\ istabip\end{tabular}}     \\ \hline
\textit{\begin{tabular}[c]{@{}c@{}}Cultural and Art \\ Activities\end{tabular}}  
                     &  \textit{9 out of 10}  & \textit{\begin{tabular}[c]{@{}c@{}}CMYLMZ, AtlasTarihDergi, \\ Siirler\_sokakta\end{tabular}}    \\ \hline
\end{tabular}
\caption{Estimated influential accounts.}
\label{influentialTable}
\end{table}
For the evaluation of the results, we performed a small survey containing 10 
people chosen among graduate students who are closely interested in
social media. We asked participants to evaluate the users with respect to their 
topic relevance and their influence on the topic. All participants 
were shown all influential account for all topics. In order to identify 
influence of a user, we asked participants to mark one of the following 
categories: \begin{enumerate*}[label=\itshape\alph*\upshape)] 
\item very influential (1), \item influential (.5), \item not influential (0)
\end{enumerate*}. Results are aggregated as average and rounded 
by $.5$ precision. We used the results of the survey to provide an evaluation of 
the selected users for the Turkish Twitter network, on a per-topic basis.

For the topic Politics, the results are very accurate for top-10. We have
observed that the dictionaries constructed for each topic has a big impact on
the results. For example, we observe that the dictionary constructed for
Politics topic contains many keywords that are related only with politics
without any ambiguity. These keywords have increased the performance of
the semantic analysis, which in turn increased the accuracy of the topic-based
network influence analysis. Top-10 list contains the president of Turkish
Republic (RT\_Erdogan), the chairman of one of the opposition parties
(kilicdarogluk), and the mayor of the capital city (06melikgokcek). It is fair
to assume that these users, who give political messages in their tweets and
who have lots of followers, should be in the top-10 influential list on
Turkish Politics topic.

The influential accounts for the Sport topic were the biggest sport clubs of
Turkey (Fenerbahce, GalatasaraySK) and one of the highest rating sport channel
(ntvspor). Their tweets were mostly related to the sport competitions, news
from clubs, etc. They have a lot of followers who actively pay attention to
what they tweet. Thus, they achieve high RT and Fav statistics, which shows
that they have a big impact on their followers. It is very reasonable that
they are the top influential accounts on this topic.

As intuitively expected, the influential accounts for the Health topic are
mostly doctor associations and governmental authorities. One of the accounts
is Republic of Turkey Ministry of Health (saglikbakanligi), which mainly
tweets about hospitals, doctors, and health regulations. Its follower numbers
can be considered as relatively high and is followed by other influential
accounts. Since its tweets have critical news potential, it has considerable
number of RTs about the health topic. The other two are doctor associations
(YYD\_tr, istabip). They are followed by many doctors, which  also have some
potential impact on the Health topic. In this topic, accurate relevance ratio
is relatively low because the constructed dictionary for this topic is not
specific enough, causing errors in semantic analyses that propagates to the
latter phase of influence estimation.

The Cultural and Art Activities topic includes users which tweet about movies,
art, books, history, etc. The top-10 influential users are perfectly matched
with the keywords. CMYLMZ is a very famous Turkish comedian, actor and producer.
He also has one of the highest follower numbers in the Turkish Twitter
network. AtlasTarihDergi is a history magazine tweeting mainly about
historical events and information which has considerable amount of followers
and RTs. The third user (Siirler\_sokakta) shares street poems and mottos, and
it's posts receive many RTs and Favs.

\subsubsection{Evaluation of Dynamic Tweet Fetching}
\noindent
We have used the same default parameter settings from the network fetching
experiments to evaluate our proposed tweet fetching methods.  
For the simplicity, we only evaluate the case of topic Politics.

\begin{figure}[t!]
\centering
\begin{subfigure}{.5\textwidth}
  \centering
  \includegraphics[width=.45\linewidth,keepaspectratio]{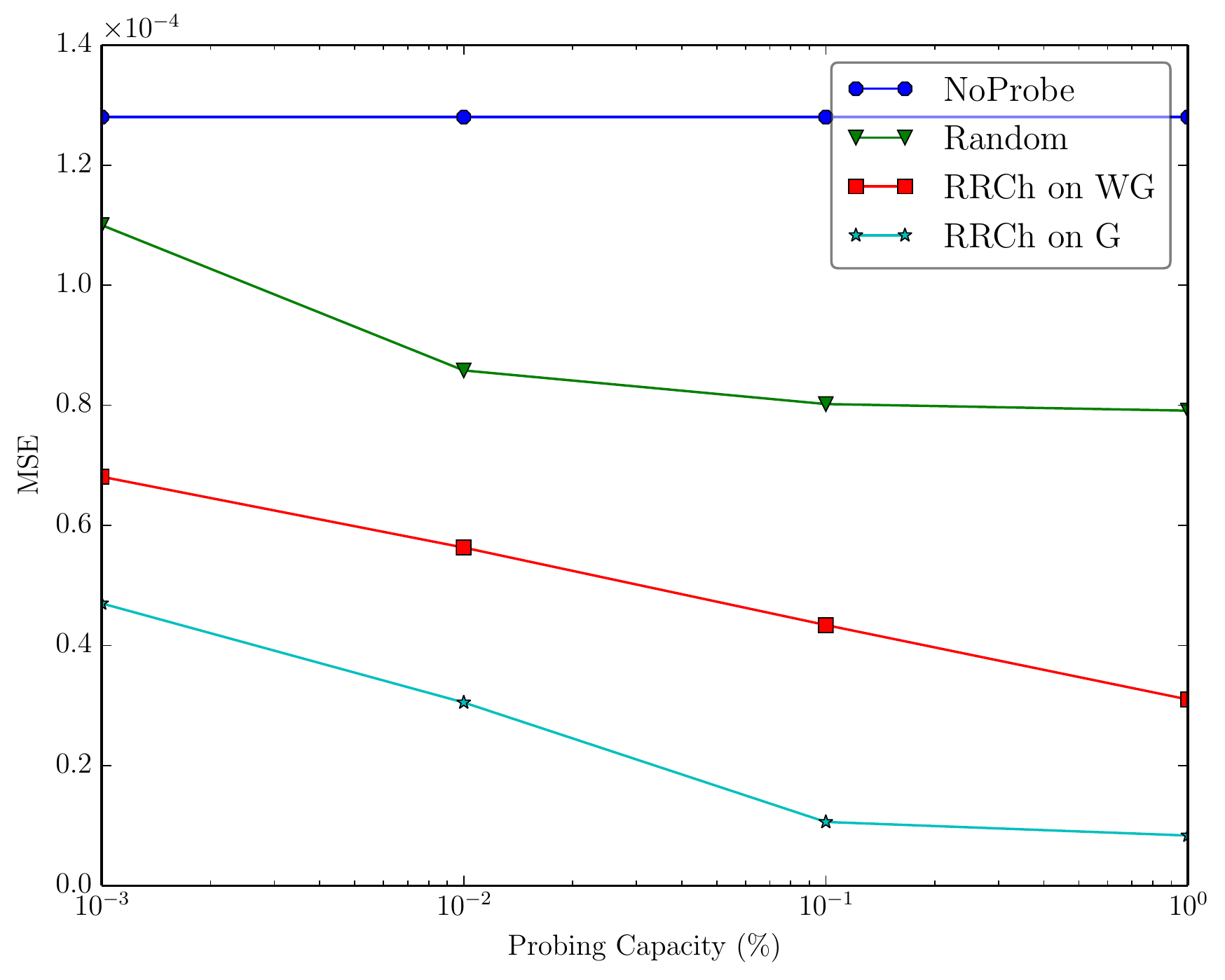}
  \caption{MSE}
  \label{fig:mseTopic}
\end{subfigure} \\
\begin{subfigure}{.5\textwidth}
  \centering
  \includegraphics[width=.75\linewidth,keepaspectratio]{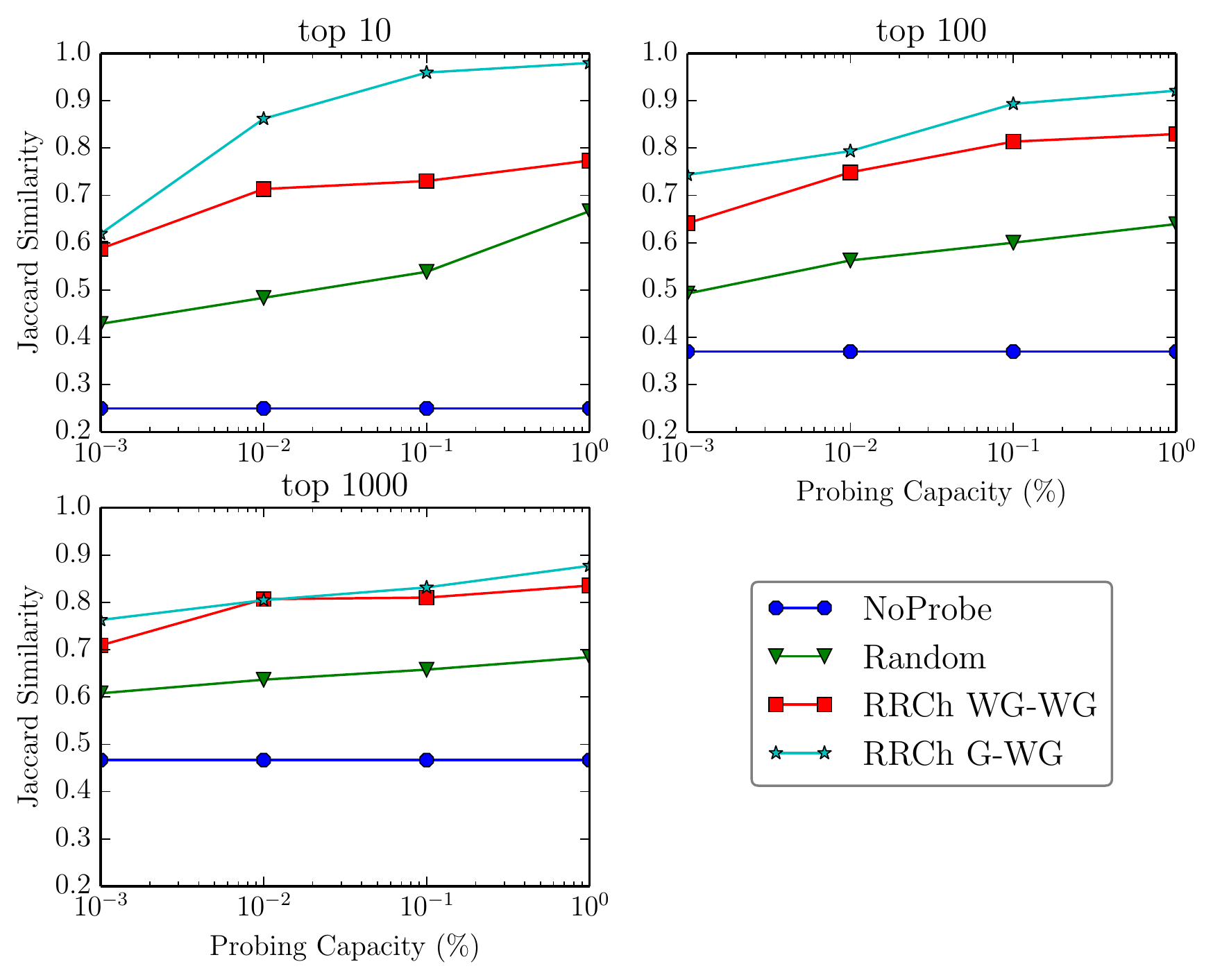}
  \caption{Jaccard similarity}
  \label{fig:jacTopic}
\end{subfigure}
\caption{Performance of Change Probing for dynamic tweet fetching.} 
\label{fig:topicPerformance}
\end{figure}

Figure~\ref{fig:topicPerformance} shows the performance of the RRCh
method for dynamic tweet fetching. For the MSE measure, global network based
$G$-$WG$ method performs $78\%$ better, and topic network based $WG$-$WG$
method performs $40\%$ better than the baselines, on average, respectively. In
Figure~\ref{fig:jacTopic}, we see that as the probing capacities increase,
$G$-$WG$ method achieves almost perfect similarity against the results
obtained using the original network, for the top-$10$ influential users. For
the top-$1000$ influential users experiment, it reaches close to $0.9$
similarity. Together with $WG$-$WG$ method, they quickly reach close to their
top performance at around $1\%$ capacity, except for the top-$10$ case. For
the latter, $WG$-$WG$ method does not enjoy the quality increase that the
$G-WG$ method enjoys with increasing capacities. When we look at the Jaccard
similarity based results, $G$-$WG$ achieves $77\%$ better and $WG$-$WG$
achieves $65\%$ better results than the baselines. Overall, the results show
us that using the globally maintained network is more advantageous.

Although $G$-$WG$ method outperforms $WG$-$WG$ method when we compare the
top-$10$ results for the two methods, they are similar in terms of the topic
relevance of their top influential users. Table~\ref{top10DynamicProbTable}
shows the topic relevance ratios for the two methods. Top-$10$ selected users
are found to be related with the topics of interest and are popular accounts
in the topic area.

\begin{table}[!ht]
\centering
\begin{tabular}{c|c|c|}
\hline
\multicolumn{1}{|c|}{\textbf{Topics}}   & \multicolumn{2}{c|}{\textbf{Topic Relevance}}      \\ \hline
\multicolumn{1}{|c|}{\textit{Politics}} & \textit{10 out of 10}    & \textit{10 out of 10}   \\ \hline
\multicolumn{1}{|c|}{\textit{Sport}}    & \textit{8 out of 10}     & \textit{9 out of 10}    \\ \hline
\multicolumn{1}{|c|}{\textit{Health}}   & \textit{5 out of 10}     & \textit{4 out of 10}    \\ \hline
\multicolumn{1}{|c|}{\textit{\begin{tabular}[c]{@{}c@{}}Cultural and Art \\ Activities\end{tabular}}}    & \textit{9 out of 10}     & \textit{9 out of 10}    \\ \hline
\multicolumn{1}{l|}{}                   & \multicolumn{1}{l|}{\textbf{$G-WG$ method}} & \multicolumn{1}{l|}{\textbf{$WG-WG$ method}} \\ \cline{2-3} 
\end{tabular}
\caption{Top-$10$ topic relevance ratios for $G$-$WG$ and $WG$-$WG$ for dynamic tweet fetching.}
\label{top10DynamicProbTable}
\end{table}

%% file: futuredirections.tex
\section{Future Directions}\label{sec:futuredirections}    
\noindent
In this section we discuss improvements and extensions to our work that are
left as future research directions.

First, the simulation technique used in this study for
evaluating the probing strategies does not take into account the following two
aspects: $i$) a snapshot of the network collected as ground truth data does
not represent an instantaneous snapshot and instead is the result of crawling,
which takes non-negligible amount of time, and $ii$) the simulation of probing
strategies assumes that the network does not change as the probing happens,
but in a real-world scenario the network can evolve during this time. A future
direction for having more accurate simulation results is to consider the
probing time explicitly as part of the simulation, while at the same time
modeling the network change as a random process.

Second, this study focuses on effectively probing the network
for capturing edge updates, which constitutes the majority of the change in
the social network.  Yet, node additions and deletions also take place in a
dynamic network. Our proposed system handles node updates by periodically
repeating the seed list construction process. We leave it as a future work to
integrate node update into the edge probing process.

Third, for the topic-based network construction, we ignore the
impact of individual tweets. We maintain a keyword corpora for user tweet sets
and perform our topic analysis over these corpora. For approximating a user's
influence on a particular topic, we scale her RT and FAV statistics with the
relative relevance of her tweet set with the given topic. This is not as
accurate as analyzing individual tweets. This is because a user may be
tweeting mostly about one topic, yet receiving most of her RTs and FAVs for
tweets posted about another topic. Integrating a topic classifier that works
at the granularity of individual tweets is left as a future work. Luckily,
such a classifier can be easily plugged into our framework. Similarly, topic
classification techniques that are more advanced than the weighted keyword
dictionaries we employed in this study can be integrated into our framework
with ease.

Last, another interesting future research direction is utilizing 
a technique that can dynamically adjust $\theta$, which controls the balance
between the last PageRank score and the change in recent PageRank scores in
Eq.~\ref{Score}. Here, one can use an adaptive value at each iteration, tuned
for each user based on some heuristic. An intelligent way of performing
adaptive $\theta$ control could potentially improve the accuracy of the
proposed techniques.

%% file: related.tex
\section{Related Work}\label{sec:related}  

\noindent Increases in the popularity of social networks and the  
availability of public data acquisition tools for them have put social
networks on the spotlight of both academic and industrial research.
Influential user estimation problem is studied by many researchers following a
wide variety of different methodologies. Within this context, some studies
introduce centrality measures in order to reflect influence of users.
\cite{wasserman1994social} introduces several definitions, such as
\textit{degree centrality}, \textit{betweenness centrality}, and
\textit{closeness centrality}. For viral marketing applications, 
\cite{domingos2001mining} develops methods for computing  
network influence from collaborative filtering databases by using heuristics
in a general descriptive probabilistic model of influence propagation.
\cite{kempe2003maximizing} addresses a similar problem by studying the linear
threshold and independent cascade models, and \cite{kempe2005influential}
presents a simple greedy algorithm for maximizing the spread of influence
using a general model of social influence, termed the decreasing cascade
model.
 
Recently, researchers have studied extracting textual information associated
with social networks. \cite{mei2008topic} studies topic modeling in social
networks and proposes a solution for text mining on the network structure.
\cite{tang2009social} introduces the topic-based social influence problem.
Their proposed model takes the result of any predefined topic modeling of a
social network and constructs a network representing topic-based influence
propagation. Distributed learning algorithms are used for this purpose, which
leverage the Map-Reduce concept. Thus, their methodology scales to large
networks. \cite{liu2010mining} combines heterogeneous links and textual
content for each user in order to mine topic-based influence. 
In another seminal work, \cite{haveliwala2002topic} studies 
topic-specific influence by using PageRank.

Another recent study~\cite{weng2010twitterrank} uses a PageRank-like measure
to find influential accounts on Twitter. They extend PageRank by using
topic-specific probabilities in the random surfer model. Although their
method is similar to ours, their influence measure utilizes the number of
posts made on a specific topic. However, this is an indirect measure that
cannot reliably capture influence. Therefore, we use topic distributions of
user posts along with their sharing statistics (re-tweets and favorites in
Twitter), which provides robust results, as it takes into account the real
impact of posts. \cite{Hong:2010:EST} conducts an empirical study of different
topic modeling strategies based on standard
\textit{Latent Dirichlet Allocation} (LDA)~\cite{blei2003latent}.
\cite{lin2011joint} proposes joint probabilistic models of influence 
and topics. Their methodology performs a topic sampling over textual contents
and tracks the topic snapshots over time. \cite{hong2011predicting} uses
re-tweets in measuring popularity and proposes machine learning techniques
to predict popularity of Twitter posts.
\cite{Szabo:2010:PPO,Yang:2011:PTV,cheng2014can} propose solutions for
predicting popularity of online content. \cite{chen2015online} studies the
topic-aware influence maximization problem. Within this context, in this work
we introduce a new method that combines topic-based analyses of posts with
their sharing popularity for the purpose of topic-based influential user
estimation.
 
Dynamic graph analysis has also attracted a lot of attention recently. In
order to maintain dynamic networks,
\cite{wolf2002optimal,Cho2003,cho2003estimating,Pandey:2005:UWC,Olston:2008:RSB} 
propose algorithms for determining web crawling schedules.
\cite{leskovec2008microscopic} studies the microscopic evolution of social
networks.
\cite{Desikan05incrementalpagerank} studies incremental PageRank on evolving
graphs. Researches have also investigated probing strategies for analyzing
evolving social networks.
\cite{bahmani2012pagerank} proposes influence proportional probing strategies
for the computation of PageRank on evolving networks and \cite{dynamicinfmax}
uses a probing strategy to capture observed image of the network by maximizing
a performance gap function. \cite{papagelis2013sampling,Valkanas:2014:MTD,nazi2014walk} study
sampling over social networks. However, these studies only focus on current
image of a network in their probing strategies. In contrast, we propose a
method which also considers evolution of the probing metrics, so that the
network could be probed more effectively.

In the context of network inference, \cite{getoor2003learning} proposes
representations for structural uncertainty and use directed graphical models
and probabilistic relational models for link structure learning. However,
their methodologies are not scalable.
\cite{ghahramani1998learning,song2009time,koskinen2007bayesian} use time
evolving graph models for social network estimation. They apply time-varying
dynamic Bayesian networks for modeling evolving network structures.
\cite{bonneau2009eight} shows that third-parties can reach a user's
information by searching a few friends. \cite{gomez2010inferring} develops a
scalable algorithm to infer influence and diffusion network based on an
assumption that all users in the network influence their neighbors with equal
probability. \cite{myers2010convexity} removes this assumption and addresses
the more general problem by formulating a maximum likelihood problem and
guarantee the optimality of the solution. \cite{Yang:2010} proposes a linear
model to predict how diffusion unfolds over time and \cite{kang2012diffusion}
proposes the notion of diffusion centrality.
\cite{yang2010modeling,rodriguez2011uncovering} studies a different problem
related to network inference. Different from these works, we use friendship
weighting method in order to infer link structures, similar to
\cite{taskar2003link,vert2004supervised,liben2007link}. However, we use
friendship weights only to infer edges between users. 
\cite{du2012learning} proposes a kernel based method and
\cite{du2013uncover} uses a continuous time model for inference. Moreover, one
can also use more informative features such as content-based influential
effects. \cite{wang2013whom} studies diffusion of tweets throughout the
Twitter network. This kind of technique could also be used in order to
estimate impact of posts.

%% file: conclusion.tex
\section{Conclusion}\label{sec:conclusion}
The rate restrictions enforced by social network service providers have a
negative impact on the third-party evolving network analysis tasks. Therefore,
we proposed probing algorithms to dynamically fetch network topology and text
data from social networks under limited probing capacities. Our proposed
solutions use the past influence trends of the users, as well as their current
influences, in order to determine the best users to probe, with the aim of
maximizing the influence estimation accuracy. In particular, we observed that
highly influential users and users with strong influence trends affect the
overall influence estimations the most. We have leveraged these two metrics
across our probing algorithms. Experimental results have shown that
considering past trends in the probing strategy increases the overall accuracy
of influence prediction. Furthermore, we improved our probing strategies by
inferring possible relations between users via link prediction algorithms. We
also developed techniques for estimating topic-based user influence in dynamic
social networks. For computing topic-based influence, we proposed methods that
consider both the place of the user in the network topology, as well as the
topic analysis performed on the user posts and the sharing statistics of these
posts. Our experimental results performed on Twitter network data has shown
improved accuracy compared to state-of-the-art methods from the literature.

%% file: acknowledgments.tex
\section{Acknowledgments}
\noindent 
Special thanks to Mr. Mehmet G{\"u}vercin for his contributions to creation of
the word dictionaries and topic relevance scoring. This work is supported in part 
by Turkish Academy of Sciences and T{\"u}rk Telekom.

%% file: bio.tex
\vspace{-1cm}
\begin{IEEEbiography}[{\vspace{-1cm}\includegraphics[width=1in,height=1.25in,clip,keepaspectratio]{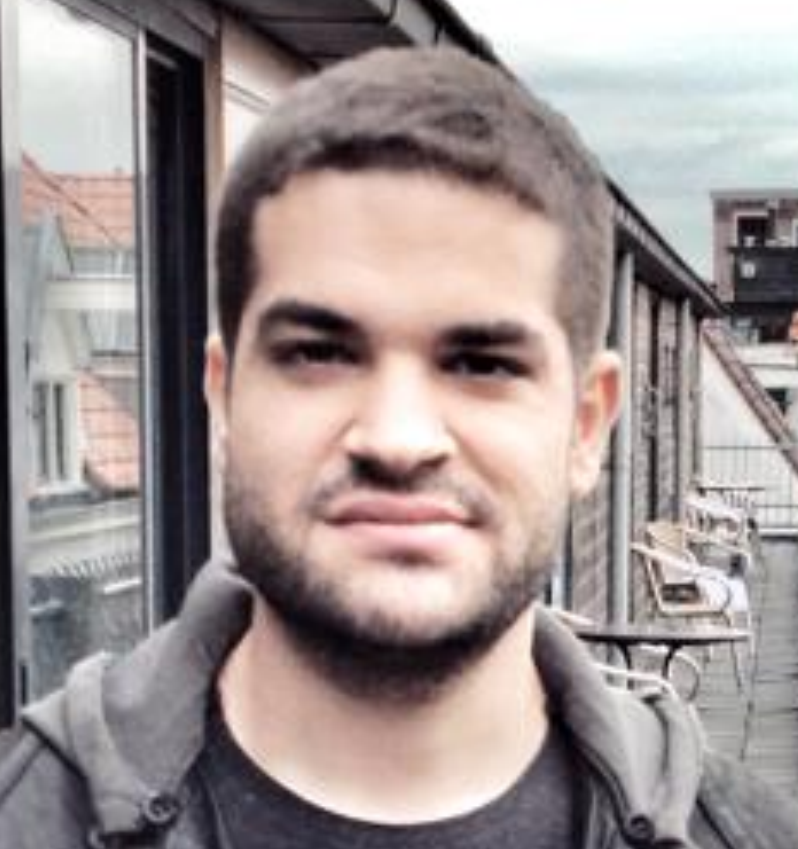}}]
{Kaan~Bing\"{o}l} is an MSc. Student in the Department of
Computer Engineering, Bilkent University, Turkey. His research interests 
are in large dynamic graphs and big data technologies.
\end{IEEEbiography}
\vspace{-1.8cm}
\begin{IEEEbiography}[{\vspace{-1cm}\includegraphics[width=1in,height=1.25in,clip,keepaspectratio]{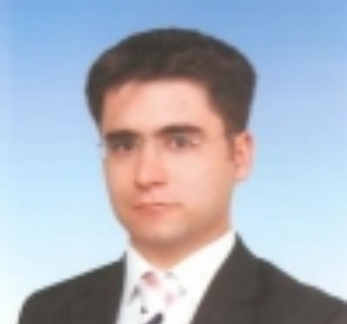}}]
{Bahaeddin~Eravc{\i}} is working towards his Ph.D. degree in computer
science at Bilkent University. His research interests are time series
data mining and data management.
\end{IEEEbiography}
\vspace{-1.8cm}
\begin{IEEEbiography}[{\vspace{-1cm}\includegraphics[width=1in,height=1.25in,clip,keepaspectratio]{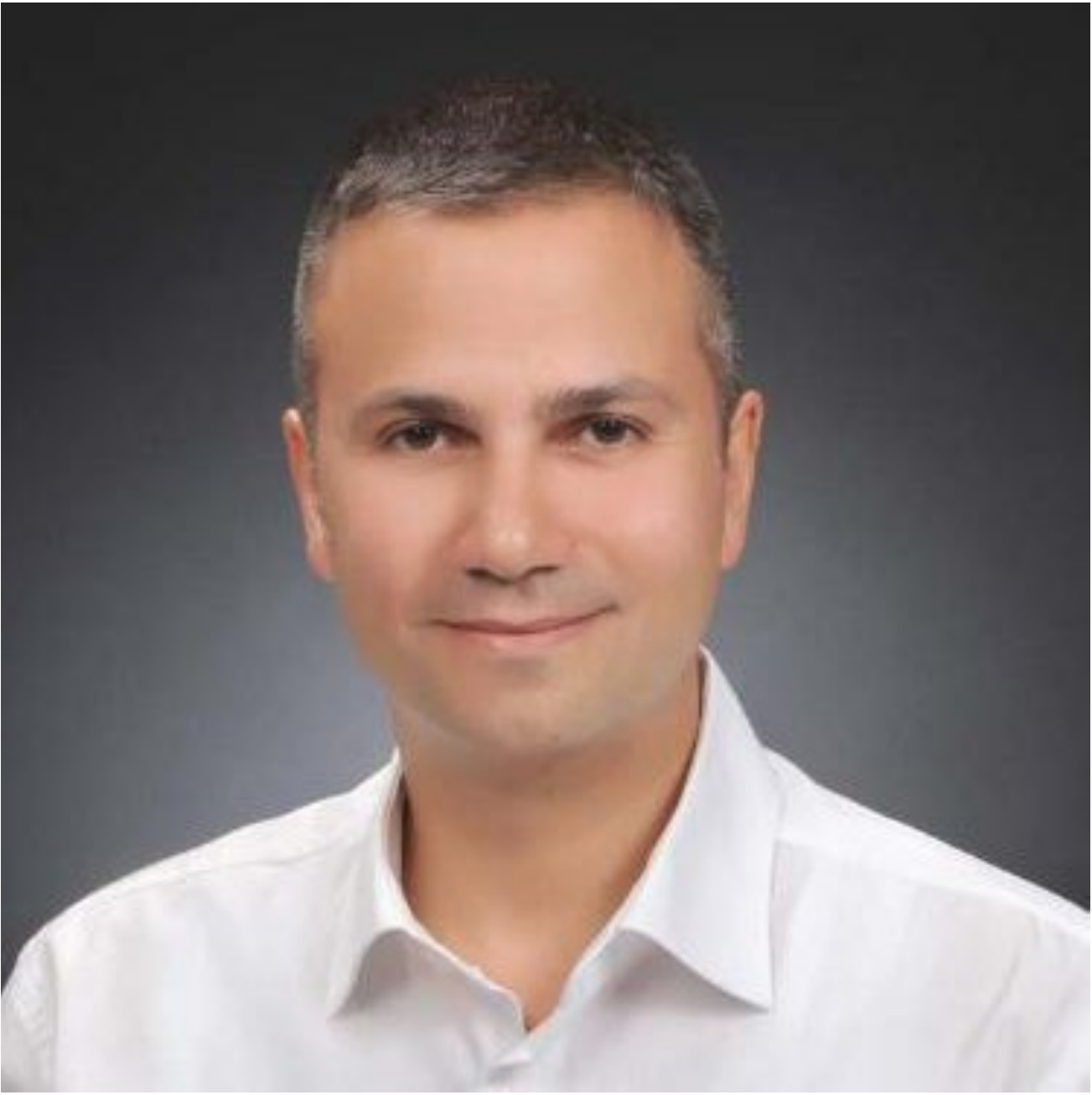}}]
{\c{C}a\u{g}r{\i}~\"{O}zgen\c{c}~Etemo\u{g}lu} is the Manager of the Big Data Software Development group at T\"urk Telekom. 
He holds a Ph.D. degree from University of California Santa Barbara, USA. His research interests are in Big Data systems.
\end{IEEEbiography}
\vspace{-1.8cm}
\begin{IEEEbiography}[{\vspace{-0.75cm}\includegraphics[width=1in,height=1.25in,clip,keepaspectratio]{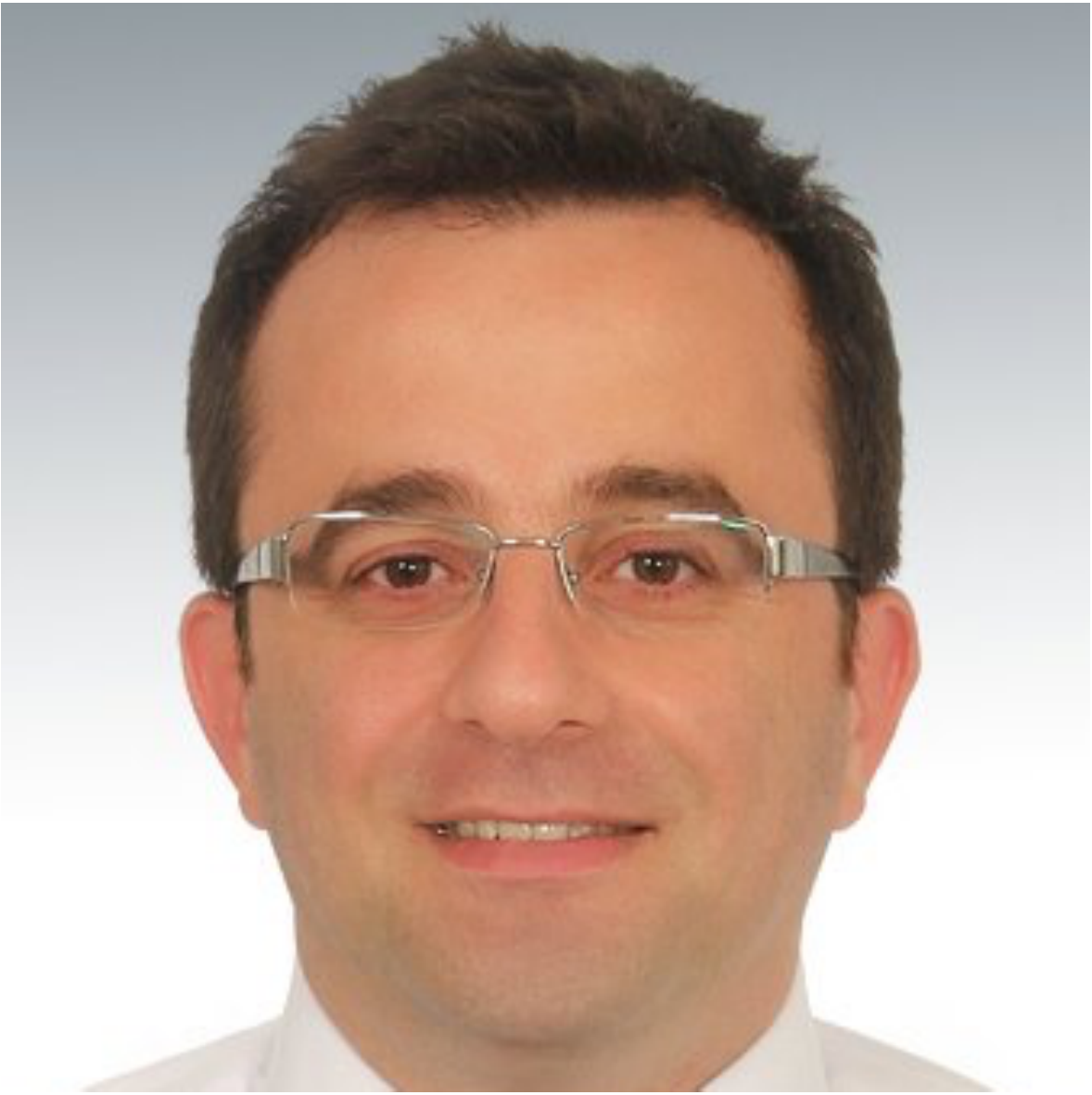}}]
{Hakan Ferhatosmano\u{g}lu} is a Professor at Bilkent University, Turkey. He was
with The Ohio State University before joining Bilkent. His current
research interests include scalable management and mining of
multi-dimensional data. He received Career awards from the US Department
of Energy, US National Science Foundation, and Turkish Academy of
Sciences. He received the Ph.D. degree in Computer Science from University
of California, Santa Barbara in 2001.
\end{IEEEbiography}
\vspace{-1.4cm}
\begin{IEEEbiography}[{\vspace{-0.75cm}\includegraphics[width=1in,height=1.25in,clip,keepaspectratio]{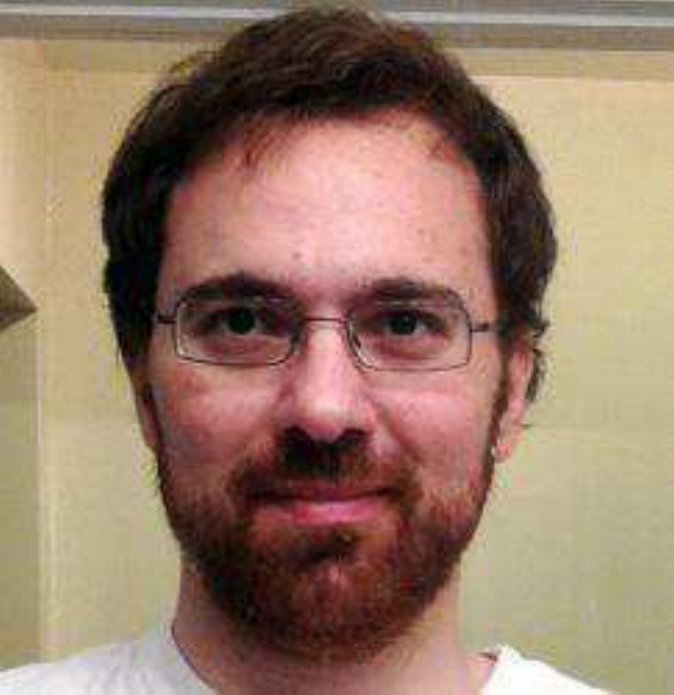}}]
{Bu\u{g}ra Gedik} is an Associate Professor in the Department of
Computer Engineering, Bilkent University, Turkey. He obtained a Ph.D. degree
in Computer Science from Georgia Tech, USA. His research interests are in
distributed data-intensive systems, with a particular focus on stream
computing and big data technologies.
\end{IEEEbiography}